\documentclass[aps,prb,twocolumn,notitlepage,superscriptaddress,showpacs]{revtex4-1}

\usepackage{amsmath,latexsym,amssymb,bm}
\usepackage{hyperref}
\usepackage{color}
\usepackage{graphicx,subfigure}
\usepackage{multirow}
\usepackage{natbib}

\begin{document}

\title{Variational wave function for an anisotropic single-hole-doped $t$-$J$ ladder}

\author{Qing-Rui Wang}
\affiliation{Institute for Advanced Study, Tsinghua University, Beijing 100084, China}
\author{Zheng Zhu}
\affiliation{Institute for Advanced Study, Tsinghua University, Beijing 100084, China}
\author{Yang Qi}
\affiliation{Institute for Advanced Study, Tsinghua University, Beijing 100084, China}
\affiliation{Perimeter Institute for Theoretical Physics, Waterloo, ON
  N2L 2Y5, Canada} 
\author{Zheng-Yu Weng}
\affiliation{Institute for Advanced Study, Tsinghua University, Beijing 100084, China}


\date{September 3, 2015}

\begin{abstract}

Based on three general guiding principles, i.e., no double occupancy constraint, accurate description of antiferromagnetism at half-filling, and the precise sign structure of the $t$-$J$ model, a new ground state wave function has been constructed recently [Weng, New J. Phys. {\bf 13}, 103039 (2011)]. In this paper, we specifically study such kind of variational ground state for the one-hole-doped anisotropic two-leg $t$-$J$ ladder using variational Monte Carlo (VMC) method. The results are then systematically compared with those recently obtained by density matrix renormalization group (DMRG) simulation. An excellent agreement is found between the VMC and DMRG results, including a ``quantum critical point'' at the anisotropy parameter $\alpha=\alpha_c\approx0.7$ (with the parameters $t/J=3$), and the emergence of charge modulation and momentum (Fermi point) reconstruction at $\alpha>\alpha_c$ due to the quantum interference of the sign structure. In particular, the wave function indicates that a Landau's quasiparticle description remains valid at $\alpha<\alpha_c$ but fails at $\alpha>\alpha_c$ due to the breakdown of the one-to-one correspondence of momentum and translational symmetry of the hole. The explicit form of the wave function provides a direct understanding on how the many-body strong correlation effect takes place non-perturbatively in a doped Mott insulator, which sheds interesting light on the two-dimensional case where the same type of wave function was proposed to describe the cuprate superconductor.

\end{abstract}


\maketitle


\section{Introduction}
\label{sec:intro}

Ground state wave function is of great importance in understanding a new state of matter. Right after the discovery of high-temperature superconductivity in the cuprate, based on the conjecture that the cuprate superconductor be a doped Mott insulator, Anderson proposed \cite{RVB1987} a resonating-valence-bond (RVB) ground state, which may be simply expressed as 
\begin{equation}
\label{BCS}
|\Psi _{\mathrm{RVB}}\rangle =\hat{P}_{\mathrm{G}}|\mathrm{BCS}\rangle,
\end{equation}
where $|\mathrm{BCS}\rangle $ denotes an ordinary BCS state and $\hat{P}_{\mathrm{G}}$ the Gutzwiller projection operator enforcing the no double occupancy constraint due to strong on-site Coulomb repulsion. Such a ground state has been intensely studied \cite{pwa_03, Lee06, gros_07} variationally since then, which will be referred to as the Anderson's one-component RVB wave function.

The no double occupancy constraint is just one of the most essential characterizations of Mott physics. The residual superexchange coupling will further cause antiferromagnetic (AF) correlations  between the singly occupied spins, leading to an AF Mott insulator at half-filling described by the Heisenberg type model. Last but not least, due to strong on-site Coulomb repulsion, the original Fermi signs of the electrons will be replaced by the so-called phase string sign structure\cite{Weng1996,Weng1997}, which has been precisely identified in the $t$-$J$\cite{Wu2008sign} and Hubbard\cite{ ZW_2014} models for arbitrary doping, temperature and dimensions on a bipartite lattice. These three constitute the basic organizing principles for the strongly correlated electrons in a doped Mott insulator.    

Based on the above three guiding principles, a new class of ground state wave function has been constructed recently, whose compact form may be written as follows\cite{Weng2011a} 
\begin{equation}
\label{scgs-0}
|\Psi _{\mathrm{G}}\rangle ={\hat{D}}^{N_h/2}|\mathrm{RVB}\rangle,
\end{equation}
where $|\mathrm{RVB}\rangle$ denotes a spin background (``vacuum'') always remaining singly occupied. The doped holes (of total number $N_h$) are created in pairs on such a ``vacuum'' via 
\begin{equation}\label{D}
\hat{D}\equiv\sum_{ij}\left[g_{ij}e^{-i\left(\hat{\Omega}_i+\hat{\Omega}_j\right)}\right]{c}_{i\uparrow }{c}_{j\downarrow },
\end{equation}
with the no double occupancy being automatically maintained. Here the sign structure is implemented by a phase shift operator
\begin{align} 
  \label{eq:PS}
  \hat{\Omega}_i \equiv \sum_l \theta_i(l)\, \hat n_{l\downarrow},
\end{align}
where $\theta_i(l)$ is a statistical angle satisfying $\theta_i(l)-\theta_l(i)=\pm \pi$ with $\hat n_{l\downarrow}$ defined as the down-spin number operator solely acting on $|\mathrm{RVB}\rangle$, commuting with the $c$-operators in $\hat{D}$.

Here the wave function $|\Psi _{\mathrm{G}}\rangle$ has a two-component RVB structure, with $|\mathrm{RVB}\rangle$ characterizing the neutral spin correlation (which reduces to the true ground state of the Heisenberg model at half-filling) and $\hat{D}$ the Cooper pairing, respectively, improving the original Anderson's one-component RVB state $|\Psi _{\mathrm{RVB}}\rangle$. The novelty of Eq.~(\ref{scgs-0}) lies in that the doped holes and the spin background are {\it nonlocally entangled} by the phase shift operator $\hat{\Omega}_i$, such that each doped hole will always feel the influence from the background spins, and vice versa. The interplay will then reshape both neutral RVB and charge pairing as a function of doping and result in a phase diagram {\it self-consistently}, which provides \cite{Weng2011a,Ma2014} a systematic understanding/explanation of AF, superconducting, and pseudogap phenomena observed in the cuprate.

In particular, if only a single hole doped into the Mott insulator is considered, $|\Psi_{\mathrm{G}}\rangle$ in Eq.~(\ref{scgs-0}) is reduced to 
\begin{align}
 \label{eq:wf}
 |\Psi_{\text{G}}\rangle_{\mathrm{1h}} = \sum_{i} {\varphi}_h(i)e^{-i\hat{\Omega}_i} {c}_{i\downarrow}|\mathrm{RVB}\rangle,
\end{align}
with ${\varphi}_h(i)$ replacing the pairing amplitude $g_{ij}$ since there is only one hole here (created by annihilating an electron with $\downarrow$-spin, without loss of generality).
Such a single-hole state should be contrasted with a more conventional Bloch type state $|\Psi_{\text{G}}\rangle_{\mathrm{1h}}\rightarrow |{\bf k}\rangle_{\mathrm{BL}}$ assuming
${\varphi}_h(i)e^{-i\hat{\Omega}_i} \propto e^{i{\bf k}\cdot {\bf r}_i}$
in Eq.~(\ref{eq:wf}). Here one has $|{\bf k}\rangle_{\mathrm{BL}} \rightarrow e^{-i{\bf k}\cdot {\bf l}} |{\bf k}\rangle_{\mathrm{BL}}$ under the translational transformation $ {\bf r}_i\rightarrow {\bf r}_i+ {\bf l} $ of the hole coordinate in a featureless spin background. But such a translational symmetry of the {\it single hole} is in general not obeyed by Eq.~(\ref{eq:wf}).

In principle, it is not {\it a priori} that the doped hole should carry a {\it conserved} momentum ${\bf k}$ in the Mott insulator, satisfying the same Bloch theorem for a doped hole in a semiconductor. If it does, then one says that the doped hole behaves like a Landau's quasiparticle with well-defined charge, spin, effective mass and momentum, which is the basis for the Fermi liquid theory of weakly interacting electrons. However, the no double occupancy constraint in a Mott insulator means that the electrons are localized at each lattice site by strong interaction at half-filling, implying the loss of the translational symmetry of the charge. A doped hole moving on the neutralized spin background of thermodynamic scale does not restore the charge translational symmetry immediately. As a matter of fact, in an early study of a single hole doped $t$-$J$ model, it has been rigorously shown \cite{Weng1996,Weng1997} that the hole acquires an irreparable many-body phase shift, i.e., the phase string, which demonstrates a general breakdown of the translational symmetry for the charge, supporting the argument of Anderson \cite{Anderson90PRL} that the hole doped into a Mott insulator does not become a well-defined quasiparticle due to a nontrivial scattering phase shift.

The two-leg $t$-$J$ ladder as a stack of two one-dimensional chains can serve as an ideal minimal model to test the novel phase string effect as well as the variational wave function Eq.~(\ref{eq:wf}), which can be accurately studied numerically by density matrix renormalization group (DMRG) method\cite{White1992}. At half-filling, the spins are short-range AF correlated such that the ground state is gapped\cite{DMRG1}. A single doped hole should not change the background spin correlation at long distance. From a more conventional point of view, the hole is expected to only carry a small distortion (spin polaron) in the spin background surrounding it with well-defined charge, spin, effective mass and a conserved momentum\cite{SCBA1,SCBA2,SCBA3,SCBA4} just like a Landau's quasiparticle. However, the DMRG study has shown exotic behaviors upon doping\cite{DMRG1,DMRG2,DMRG3,DMRG4}. Specifically, by tuning an anisotropic parameter $\alpha$ of the two-leg ladder, a critical value at $\alpha_c$ is found\cite{DMRG3,DMRG4} such that in the strong rung case of $\alpha<\alpha_c$ the doped hole indeed behaves like a conventional quasiparticle with a well-defined momentum. However, at $\alpha>\alpha_c$ the momentum splits continuously as a function of $\alpha-\alpha_c$, accompanied by incommensurate charge modulations which violate the translational symmetry\cite{DMRG3,DMRG4}. By examining\cite{DMRG1,DMRG3} the charge response to an inserting flux into the ring of the two-leg ladder, an exponential decay with the circumference of the ring indicates the doped charge loses its phase coherence or momentum conservation to become ``localized'' at a sufficiently long distance at $\alpha>\alpha_c$, in sharp contrast to a coherent quasiparticle behavior at $\alpha<\alpha_c$. Further surprising arises\cite{DMRG2,DMRG3} when two holes are injected into the gapped spin ladder, where a strong binding between the two holes occurs at $\alpha>\alpha_c$ and simultaneously the charge modulation disappears with restoring the translational symmetry.

Microscopically, the above novel properties can all be attributed to the phase string sign structure hidden in the bipartite $t$-$J$ ladder. As has been clearly demonstrated in the DMRG calculations\cite{DMRG1,DMRG2,DMRG3,DMRG4}, if one artificially turns off the phase string sign structure in the $t$-$J$ model, which results in the so-called $\sigma$$\cdot t$-$J$ model\cite{DMRG1}, the ordinary Bloch wave behavior of the doped hole is immediately recovered in the {\it whole} regime of $\alpha$ with no more critical $\alpha_c$. Simultaneously two doped holes are no longer paired\cite{DMRG3}. Although how the phase string effect is responsible for the physics at $\alpha>\alpha_c$ has been qualitatively discussed in Ref.~\onlinecite{DMRG4}, a microscopic and quantitative understanding of the DMRG results is still lacking. 

Recently, a paper by White, Scalapino, and Kivelson \cite{Kivelson} has reconfirmed the existence of $\alpha_c$ in the one-hole-doped two-leg ladder by DMRG under open boundary condition, together with the charge modulation and the momentum splitting at $\alpha>\alpha_c$ found in the earlier works. But they gave a different physical interpretation on the nature of the hole state at $\alpha>\alpha_c$ and argued that the hole would still behave like a Bloch quasiparticle at momenta different from that at $\alpha<\alpha_c$. However, many issues remain unanswered there, including the microscopic origin of the critical $\alpha_c$, the necessary role of the phase string effect that causes the charge modulation and momentum splitting in contrast to the $\sigma$$\cdot t$-$J$ model, the pairing between two doped holes, as well as the oscillation and exponential decay of the energy difference with the ladder length under a periodic boundary condition with inserting different fluxes, etc. In particular, how to meaningfully identify the Landau type quasiparticle is actually rather subtle in such a strongly correlated system, as to be clearly shown in this work.

In this paper, we study the one-hole-doped two-leg $t$-$J$ ladder based on the ground state $|\Psi_{\text{G}}\rangle_{\mathrm{1h}}$ [Eq.~(\ref{eq:wf})] using variational Monte Carlo (VMC) method. The continuous phase transition at $\alpha_c$ will be naturally reproduced by $|\Psi_{\text{G}}\rangle_{\mathrm{1h}}$ as a function of $\alpha$. As a matter of fact, we find that $\alpha_c\approx0.7$ at the $t/J=3$ matches with that found by DMRG\cite{DMRG3,DMRG4,Kivelson} {\it quantitatively}. Such a wave function can then provide a direct understanding on how a doped hole moves on a Mott-insulator spin background with short-ranged AF correlations. 

At small $\alpha$ ($<\alpha_c$), the ground state $|\Psi_{\text{G}}\rangle_{\mathrm{1h}}$ is shown to have a finite overlap with the simple Bloch wave state $|{\bf k}\rangle_{\mathrm{BL}}$ at $\mathbf k=\mathbf k_0=(\pi,0)$. Conversely, if the Bloch wave function $|{\bf k}\rangle_{\mathrm{BL}}$ is used as the variational state, the same $\mathbf k_0$ is also reproduced. It means that the one-hole state $|\Psi_{\text{G}}\rangle_{\mathrm{1h}}$ indeed describes a coherent quasiparticle of the Landau's paradigm. Namely, the quasiparticle has the same quantum numbers (charge, spin, and momentum) in both states, which differ only by the effective mass and thus allow for an adiabatic connection between them. In fact, in the limit of $\alpha\rightarrow 0$, the phase factor $e^{-\hat\Omega_i}$ in Eq.~(\ref{eq:wf}) can be absorbed into the hole wave function $\varphi_h(i)$ such that $|\Psi_{\text{G}}\rangle_{\mathrm{1h}}$ is explicitly shown to be smoothly connected to the Bloch state $|{\bf k}\rangle_{\mathrm{BL}}$.

At $\alpha>\alpha_c$, an incommensurate momentum splitting (or Fermi point reconstruction) is exhibited in $|\Psi_{\text{G}}\rangle_{\mathrm{1h}}$, accompanying a charge density modulation, also in excellent agreement with the DMRG result. However, if one still uses the Bloch state $|{\bf k}\rangle_{\mathrm{BL}}$ as a variational state, the momentum carried by the hole is found to always remain commensurate at $\mathbf k_0=(0,0)$, such that there is no overlap between $|\Psi_{\text{G}}\rangle_{\mathrm{1h}}$ and $|{\bf k}_0\rangle_{\mathrm{BL}}$ at $\alpha>\alpha_c$. Consequently the doped hole can no longer be described as a Landau-type quasiparticle due to the breakdown of the one-to-one correspondence between a bare electron and a quasiparticle with the same momentum.

Here the incommensurate momenta in $|\Psi_{\text{G}}\rangle_{\mathrm{1h}}$ as well as the charge density modulation can be clearly related to the intrinsic quantum interference pattern due to the phase string effect. The results are in sharp contrast to the scenario in which the charge modulation is simply interpreted\cite{Kivelson} as a standing wave of two opposite-propagating Bloch waves mixed only by the reflection at {\it open boundaries} of the two-leg ladder. In other words, the present charge modulation is related to the absence of the translational symmetry in the bulk of the ladder where the phase string effect is unscreened at $\alpha>\alpha_c$ because of the spatial separation of the hole and its spin partner with increasing $\alpha$\cite{DMRG3,DMRG4}. On the other hand, in the $\sigma\cdot$$t$-$J$ model where the phase string sign structure is absent, we show that the single-hole ground state indeed reduces to the Bloch-like one $|{\bf k}_0\rangle_{\mathrm{BL}}$ with ${\bf k}_0=(0,0)$, in which the Landau's quasiparticle description works in the whole regime of $0<\alpha<1$, again in good agreement with the DMRG result.

The bottomline is that the ground state $|\Psi_{\text{G}}\rangle_{\mathrm{1h}}$ can well capture the essential properties of the one-hole-doped two-leg $t$-$J$ ladder found in the DMRG simulation. Given the gapped spin vacuum, these novel properties are solely associated with the phase string induced by the doped hole, whose effect cannot be reduced to simply renormalizing the effective mass of a Landau's quasiparticle at $\alpha>\alpha_c$. Here the many-body phase string factor $\hat{\Omega}_i$ in Eq.~(\ref{eq:wf}) prohibits a perturbative approach starting from a Bloch state, leading to the intrinsic translational symmetry breaking of the charge. In the end of the paper, we shall also briefly discuss how the phase string effect further renders the hole self-localized through an ultimate translational symmetry breaking, which involves a many-body correction to the hole wave function $\varphi_h(i)$.

The rest of this paper is organized as follows. In Sec.~\ref{sec:wf}, we introduce the model and further discuss the Bloch-like and non-Bloch-like single-hole variational wave functions. The procedure of the VMC for determining the variational parameters of the wave functions is then outlined, with the details presented in Appendix~\ref{Appen:MC_for_1h_wf}. In Sec.~\ref{sec:transition}, we identify a second order phase transition at $\alpha_c$. The changes of physical properties, such as charge density modulation, momentum distribution and quasiparticle weight, are investigated by VMC and compared with the DMRG results. Thereafter, the physical nature of the variational wave function is further examined. As a comparison, we then show both analytically and numerically that the Bloch-like wave function well captures the essential properties of the $\sigma$$\cdot$$t$-$J$ model, of which the sign structure is merely the Marshall sign\cite{Marshall1955} rather than the phase string (see Appendix~\ref{Appen:sigma}). Finally Sec.~\ref{sec:conclusion} is devoted to the conclusion and discussion.

\section{Variational Approach}
\label{sec:wf}

\subsection{The model}

In this work, we focus on an anisotropic $t$-$J$ model on a two-leg square lattice ladder \cite{ladder1,ladder2,ladder3,ladder4,ladder5,ladder6,ladder7,ladder8,ladder9,ladder10,ladder11,ladder12,ladder13,DMRG3,DMRG4}:
\begin{align}
  \label{eq:HtJ}
  H_{tJ} &= \sum_{\langle ij\rangle} \left(H_{ij}^t + H_{ij}^J\right), \\
  \label{eq:Ht}
  H_{ij}^t &= -  \alpha_{ij} t\sum_{\sigma}\ c_{i\sigma}^\dagger c_{j\sigma} + \mathrm{h.c.}, \\
  \label{eq:HJ}
  H_{ij}^J &= \alpha_{ij} J \left( {\mathbf S}_i\cdot{\mathbf S}_j -\frac{1}{4}n_i n_j\right),
\end{align}
where $\alpha_{ij}=\alpha$ ($\alpha_{ij}=1$) if the nearest-neighbor bond $\langle ij \rangle$ is parallel (perpendicular) to the chain direction. The dimensionless parameter $\alpha$ controls the anisotropy of the model. We fix $t/J=3$ in this paper for simplicity. Here ${\mathbf S}_i$ and $n_i$ are the electron spin and number operators at site $i$, respectively, with
the no double occupancy constraint $n_i\leq 1$ always enforced on the Hilbert space.

We shall also discuss the so-called $\sigma$$\cdot$$t$-$J$ model \cite{DMRG1,DMRG2,DMRG3,DMRG4} on a two-leg ladder lattice, which differs from the $t$-$J$ model only by replacing the hopping term $H_{ij}^t$ in Eq.~(\ref{eq:Ht}) with
\begin{align}
  \label{eq:sHt}
  H_{ij}^{\sigma\cdot t} &= - \alpha_{ij} t\sum_{\sigma}\sigma c_{i\sigma}^\dagger c_{j\sigma} + \mathrm{h.c.},
  \end{align}
where a spin-dependent sign $\sigma=\pm 1$ is added to each step of hole hopping. A comparative study of these two models will reveal the fundamental physics hidden in the Mott physics.

\subsection{Ground state at half-filling}
\label{sec:RVB}

At half-filling, the two-leg $t$-$J$ ladder reduces to a pure Heisenberg spin ladder, whose ground state $|\mathrm{RVB}\rangle$ is AF short-range-correlated, separated by a finite energy gap from the first excited state\cite{DMRG1,DMRG2}. As a good starting point, Liang-Doucot-Anderson type \cite{LDA1988} bosonic RVB variational wave function will provide an excellent description:
\begin{align}
  \label{eq:RVB1}
  |\mathrm{RVB}\rangle = \sum_{v}w_v |v\rangle.
\end{align}
Here $|v\rangle=\sum_{\{\sigma\}} \left(\prod_{(ij)\in v} \epsilon_{\sigma_i,\sigma_j}\right) c_{1\sigma_1}^\dagger ... c_{N\sigma_N}^\dagger |0\rangle$ is a singlet pairing valence bond (VB) state specified by the dimer covering configuration $v$ \footnote{\label{ftn:VB} Here we express the ``bosonic'' VB states by electron creation operators.}. The antisymmetric Levi-Civita symbol $\epsilon_{\sigma_i,\sigma_j}$ ensures the singlet paring between spins on sites $i$ and $j$. The amplitude of each VB state $|v\rangle$ is factorized by $w_v=\prod_{(ij)\in v} h_{ij}$, where $h_{ij}$ is a \emph{non-negative} function depending on sites $i$ and $j$ belonging to opposite sublattices, respectively. Such a factorization tremendously decreases the number of variational parameters, with $h_{ij}$'s chosen such that $\langle\mathrm{RVB}|\mathrm{RVB}\rangle=1$. Moreover, the variational wave function Eq.~(\ref{eq:RVB1}) satisfies the exact \emph{Marshall sign rule} for bipartite Heisenberg models \cite{Marshall1955}.

\begin{figure}[h]
\includegraphics[width=7cm,height=11.2cm]{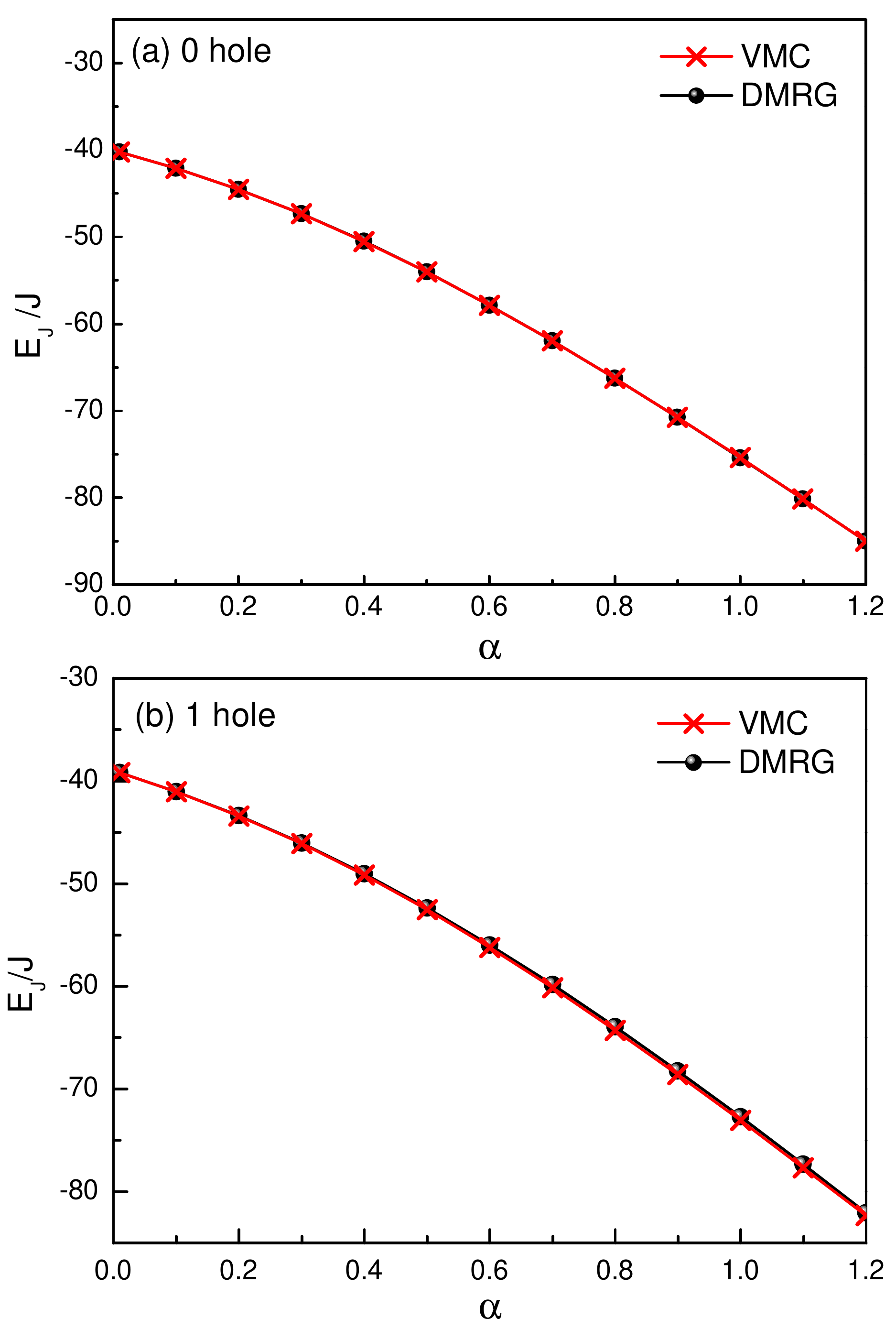}
\caption{\label{fig:EJ0h}(color online). The superexchange energy $E_J$ shows no singularity as a function of the anisotropic parameter $\alpha$. (a) The VMC calculation for the state $|\mathrm{RVB}\rangle$ in Eq.~(\ref{eq:RVB1}) (red crosses) and the DMRG result (black dots) in the pure two-leg spin ladder; (b) The superexchange energy of the single-hole-doped variational state Eq.~(\ref{eq:wf}) and the DMRG result. The two-leg ladder has size $N=N_x\times N_y=40\times2$ under the open boundary condition.}
\end{figure}

By using the wave function $|\mathrm{RVB}\rangle$ in Eq.~(\ref{eq:RVB1}), we perform the VMC calculation \cite{LDA1988} of the superexchange Heisenberg model Eq.~(\ref{eq:HJ}) on a lattice with size $N=N_x\times N_y=40\times2$ under an open boundary condition. The optimized superexchange energies $E_J$ calculated by VMC are in an excellent agreement with the DMRG energies for different $\alpha$ as shown in Fig.~\ref{fig:EJ0h} (a).

\subsection{Single-hole-doped variational state}

Once the half-filling ground state $|\mathrm{RVB}\rangle$ is accurately known, the ground state for the single hole case can be generically constructed by removing an electron (say, with a $\downarrow$ spin) from the vacuum state $|\mathrm{RVB}\rangle$ as follows
\begin{align}
 \label{eq:wf0}
  |\Psi_{\text{G}}\rangle_{\mathrm{1h}} = \sum_{i} \hat{\Phi}_h(i)\, {c}_{i\downarrow}|\mathrm{RVB}\rangle ~.
\end{align}
The wave function $\hat{\Phi}_h(i)$ is a many-body operator which involves the hole coordinate $i$ and at the same time can be spin-dependent. Generally the latter accounts for the spin background {\it response} to the creation of a bare hole at site $i$, $c_{i\downarrow}|\mathrm{RVB}\rangle$. It is usually called the ``spin-polaron'' effect\cite{SCBA1,SCBA2,SCBA3,SCBA4}. A quasiparticle description is valid if such an effect remains featureless, meaning that it only renormalizes the effective mass and the wave function spectral weight without changing the momentum. In particular, in the present two-leg ladder case, spins are gapped at half-filling over the whole range of $0\leq\alpha<\infty$ such that the correction of the spin-polaron effect to the renormalization is expected to be weak. Thus, in the present work we shall always neglect a \emph{featureless} spin-polaronic correction to $\hat{\Phi}_h(i)$, although it may be still important to improve the variational energy.

\subsubsection{A Bloch-like state}

Then, if the {\it whole} spin-polaron correction to the bare hole ``Wannier basis'' $c_{i\downarrow}|\mathrm{RVB}\rangle$ is neglected, $\hat{\Phi}_h(i)$ will reduce to a single-particle Bloch wave function
\begin{align}
  \label{eq:wf10}
  \hat{\Phi}_h(i)= \sqrt{\frac{2}{N}}\, e^{i{\bf k}\cdot {\bf r}_i},
\end{align}
by assuming a translational symmetry for the hole (which is not \emph{a priori} in a many-body system, see below). Correspondingly $|\Psi_{\text{G}} \rangle_{\mathrm{1h}} \equiv |{\bf k}\rangle_{\text{BL}}$ is uniquely specified by the momentum ${\bf k}$ of the hole without involving any other variational parameters:
\begin{align}
 \label{eq:wf1}
 |{\bf k}\rangle_{\text{BL}} \equiv \sqrt{\frac{2}{N}} \sum_{i}e^{i{\bf k}\cdot {\bf r}_i}{c}_{i\downarrow}|\mathrm{RVB}\rangle.
\end{align}

\subsubsection{A non-Bloch-like state}
\label{sec:nonBLcon}

Nevertheless, even if the \emph{longitudinal} (amplitude) spin-polaronic effect is negligible in a spin gapped background, a \emph{transverse} or many-body phase shift of the spin background in response to the creation of the bare hole may still play a crucial role in the present strongly correlated system. Specifically, one may construct a new variational wave function as given in Eq.~(\ref{eq:wf}):
\begin{align}
 \label{eq:wf2}
 \hat{\Phi}_h(i) = \varphi_h(i) e^{-i\hat\Omega_i},
\end{align}
where the phase factor $e^{-i\hat\Omega_i}$, defined in Eq.~(\ref{eq:PS}), is a nonlocal operator depending on the spin configuration in the vacuum. [The normalization $\langle\Psi_\mathrm G|\Psi_\mathrm G\rangle=1$ implies the normalization of the hole wave function $\sum_i|\varphi_h(i)|^2=2$.
] Note that the new ``Wannier basis'' $\tilde{c}_{i\downarrow}|\mathrm{RVB}\rangle \equiv e^{-i\hat\Omega_i}c_{i\downarrow}|\mathrm{RVB}\rangle$ still remains invariant under the \emph{whole} hole-spin translational operation. But $\varphi_h(i)$, determined variationally as a \emph{single-hole} wave function, is no longer necessarily Bloch-wave-like as in Eq.~(\ref{eq:wf1}). 

It is important to point out that, in contrast to a conventional weakly-interacting system, a Bloch wave construction is not automatically valid for a strongly correlated many-body system. Here one actually deals with a doped hole moving in a quantum spin vacuum rather than an inertia translationally invariant vacuum, say, in a semiconductor. In the former, the translational symmetry for the charge is not upheld generally for a \emph{relative} motion with regard to the charge neutral (Mott insulator ) spin background. (Note that this does not contradict to the translational symmetry of the \emph{total} system composed of the charge and spins as a whole.) As a matter of fact, it was rigorously shown \cite{Weng1996, Weng1997,Wu2008sign} that a hole transverses along a closed path $c$ in a doped Mott insulator, described by the $t$-$J$ model on a bipartite lattice, will always pick up a nontrivial phase string factor $(-1)^{N^h_{\downarrow} (c)}$, where $N^h_{\downarrow} (c)$ denotes the total number of down spins exchanged with the hole along the path. Clearly $(-1)^{N^h_{\downarrow} (c)}$ represents a non-integrable (path-dependent or Berry-like) phase factor associated with the motion of the doped hole, which generally breaks the translational symmetry.

In the present single-hole-doped two-leg ladder, one may expect that the phase string effect be strongly reduced over a long-wavelength scale due to the presence of an energy gap in the spin background, in contrast to a gapless case. However, due to the singular and nonlocal nature of the phase string picked up by the doped hole, it is still very crucial to carefully treat such an effect in an energetic variational procedure, which involves the nearest-neighbor hopping and superexchange processes where the quantum interference of the phase strings plays a critical role.

The phase-string operator $\hat\Omega_i$ in Eq.~(\ref{eq:PS}) can produce a phase shift $\pm \pi$ each time when the hole and a down-spin exchange positions during the hopping. In this way, the above-mentioned singular phase string $(-1)^{N^h_{\downarrow} (c)}$ gets accurately encoded by $e^{-i\hat\Omega_i}$ in Eq.~(\ref{eq:wf}). Consequently $\varphi_h(i)$ becomes a much smoother wave function which can be then determined variationally. In this sense, the phase-string factor $e^{-i\hat{\Omega_i}}$ regulates the singular phase string effect in the $t$-$J$ model and transforms the model into a perturbative-treatable formalism (its version at arbitrary doping [cf. Eq. (\ref{scgs-0})] has been previously obtained in Ref.~\onlinecite{Weng2011a}). Here it is instructive to point out that the topological phase string factor $e^{-i\hat{\Omega_i}}$ enforces the mutual statistics \cite{Zaanen_2011,Weng2011b} between the doped hole and the down-spins as indicated by the above sign structure. It plays the same role as the statistical phase factor $\prod_{i<j}(z_i-z_j)^3$  in the Laughlin wave function for $\nu=1/3$ fractional quantum Hall system \cite{Laughlin1983}, which ensures the anyonic statistics of the same species rather than the two different species as in the present case \cite{Zaanen_2011,Weng2011b}. In both cases, a traditional perturbative analysis is applicable only after explicitly identifying the topological/statistical phase factor.

The statistical angle $\theta_i(l)$ may have different choices. A natural and symmetric choice of $\theta_i(l)$ in two-dimensions is $\theta_i(l)=\theta^0_i(l)\equiv \mathrm{Im} \ln (z_i-z_l)$, where $z_i=x_i+iy_i$ is the complex coordinate. In one-dimension, it further reduces to $\theta_i(l)=\pm \pi \theta(i-l) $ with $\theta(i-l)=1$ or $0$ at $i>l$ or $i<l$, respectively. In the present anisotropic two-leg ladder, one may introduce a variational parameter $\lambda$ ($0\leq\lambda\leq\infty$):
\begin{align}
\label{eq:theta}
\tan\theta_i(l) = \lambda\,\tan\theta_i^0(l)~,
\end{align}
where $e^{i\theta_i(l)}$ and $e^{i\theta_i^0(l)}$ are located in the same quadrant of the complex plane. In the following we shall see that $\lambda \rightarrow 0$ in the strong rung regime $\alpha<\alpha_c$, while $\lambda \rightarrow \infty $ in the decouple chain limit $\alpha \rightarrow \infty$.

\subsection{Variational procedure} 
\label{sec:Emin}

Based on the variational wave functions given in Eqs.~(\ref{eq:wf}) and (\ref{eq:wf1}), one can decide the ground state by optimizing the total energy via a VMC procedure outlined as follows.

(1) Firstly, the half-filling ground state $|\mathrm{RVB}\rangle$ in Eq.~(\ref{eq:RVB1}) is optimized as discussed in Sec.~\ref{sec:RVB} [cf. Fig.~\ref{fig:EJ0h} (a)]. Upon doping one hole into the two-leg spin ladder, the variational parameters $h_{ij}$'s should remain unchanged in the thermodynamic limit.

(2) Based the Bloch-like wave function Eq.~(\ref{eq:wf1}), one finds that the hopping energy is given by
\begin{equation}\label{eq:etBL}
E_t\equiv\langle H_t \rangle_{\text{BL}} =2t^x_{\text{BL}}\cos k_x+t^y_{\text{BL}}\cos k_y,
\end{equation}
with $t^{x, y}_{\text{BL}}=t(1+4\langle\mathrm{RVB}|{\bf S}_i\cdot{\bf S}_j|\mathrm{RVB}\rangle)/4 $ ($ij$ are the nearest neighbors). From Eq.~(\ref{eq:etBL}), one sees that the only variational parameter is the momentum ${\bf k}$ which minimizes the hopping energy at ${\bf k}_0=(\pi, 0)$ if $t^x_{\text{BL}}>0$ and ${\bf k}_0=(0, 0)$ if $t^x_{\text{BL}}<0$ (with $t^y_{\text{BL}}<0$).

(3) On the other hand, based on the non-Bloch-like wave function Eq.~(\ref{eq:wf}), 
\begin{equation}\label{eq:et}
E_t\equiv \langle H_t \rangle = - \sum_{\langle ij \rangle} \tilde{t}_{ij} \varphi_h^\ast(j) \varphi_h(i) + \mathrm{h.c.},
\end{equation}
in which the hopping matrix element $\tilde{t}_{ij}$ is given by
\begin{equation}\label{tildet}
\tilde{t}_{ij}\equiv -\alpha_{ij} \sum_{\sigma}\langle\mathrm{RVB}|c^{\dagger}_{j\downarrow}c_{j\sigma}e^{i\hat{\Omega}_j-i\hat{\Omega}_i}c_{i\sigma}^\dagger {c}_{i\downarrow}|\mathrm{RVB}\rangle~,
\end{equation}
which can be directly computed in the variational calculation (see Appendix~\ref{Appen:MC_Ht}).
One then determines $\varphi_h(i)$ by diagonalizing Eq.~(\ref{eq:et}) under a given $\lambda$. Finally, the total energy $E_{\mathrm{tot}}\equiv E_t+E_J$ is minimized by optimizing $\lambda$.

We shall carry out the above-sketched variational procedure by using the loop update Monte Carlo algorithm \cite{Sandvik2010}. The detail formulas used in computing the above superexchange and hopping energies are given in Appendix~\ref{Appen:MC_HJ} and \ref{Appen:MC_Ht}, respectively.

\section{Ground state properties of the two-leg ladder doped by one hole}
\label{sec:transition}

Based on the VMC calculation outlined in the previous section, we present the ground state properties of $|\Psi_\mathrm G\rangle_{\mathrm{1h}} $ in Eq.~(\ref{eq:wf}) below, in comparison with the DMRG simulations as well as the conventional Bloch state $|{\bf k}\rangle_{\mathrm{BL}}$ satisfying the translational symmetry.

\subsection{``Quantum critical point'' at $\alpha=\alpha_c$}

The ground state energy of the two-leg $t$-$J$ ladder, $E_{\mathrm{tot}}$, is composed of the hopping energy $E_t$ and the superexchange energy $E_J$. As the starting point at half-filling, $|\mathrm{RVB}\rangle$ gives rise to an excellent energy $E_J$ in comparison with the DMRG result as shown in Fig.~\ref{fig:EJ0h} (a). $E_J$ remains a smooth function of $\alpha$ for $|\Psi_\mathrm G\rangle_{\mathrm{1h}} $ upon one-hole-doping, which is shown in Fig.~\ref{fig:EJ0h} (b) together with the DMRG data.

\begin{figure}[h]
\includegraphics[width=8cm,height=6cm]{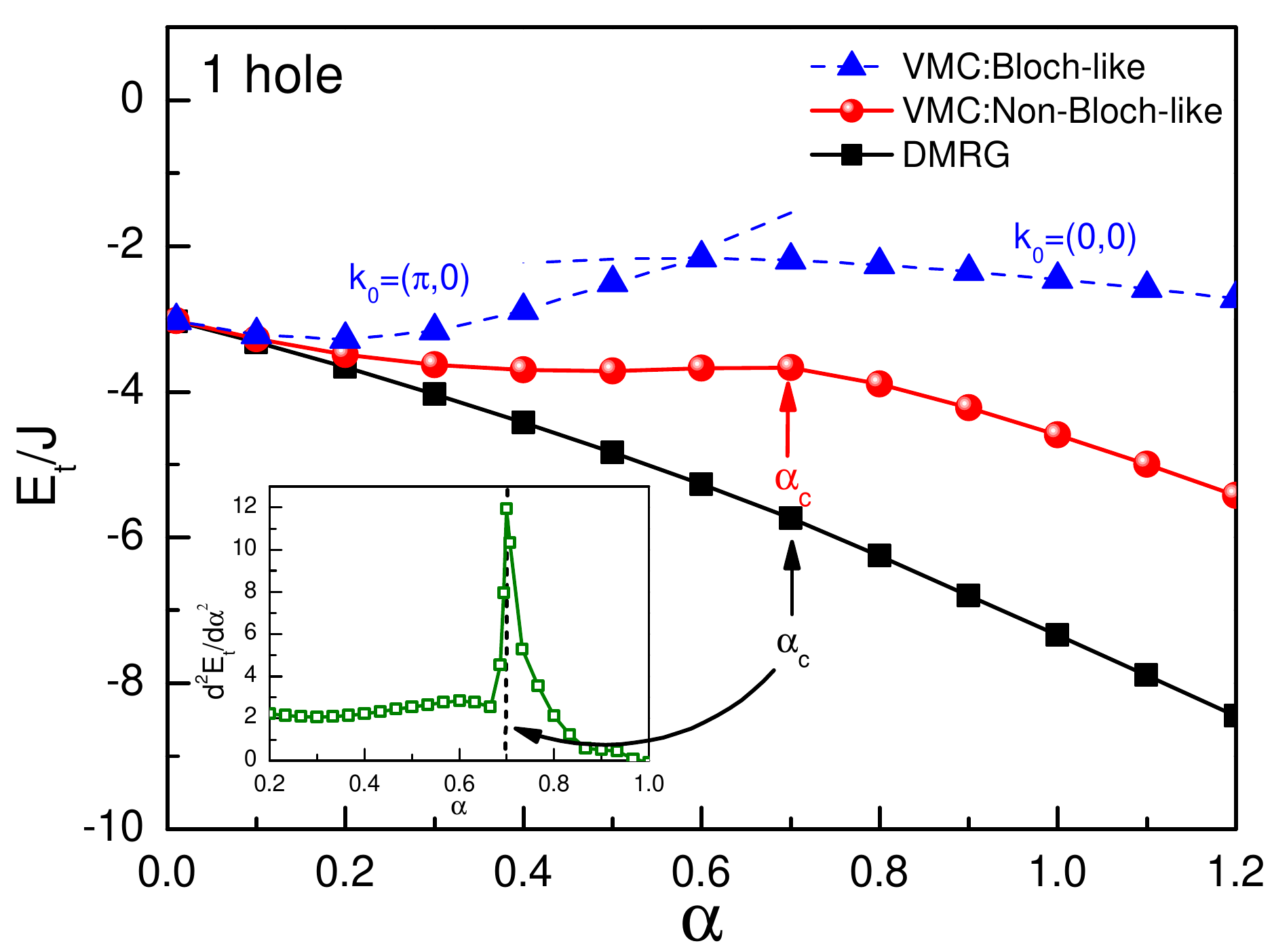}
\caption{\label{fig:Et}(color online). Hopping energies $E_t(\alpha)$ for the non-Bloch-like wave function Eq.~(\ref{eq:wf}), the DMRG ground state\cite{DMRG3}, and the Bloch-like wave function Eq.~(\ref{eq:wf1}), respectively. The inset shows the second order derivative of the DMRG energy, indicating a second order transition takes place at $\alpha_c\approx0.7$. A similar critical point of $\alpha_c\approx0.7$ is also identified for the non-Bloch-like wave function Eq.~(\ref{eq:wf}). By contrast, there is a level crossing at $\alpha_c^\mathrm{BL}\approx0.6$ for the Bloch-like wave function Eq.~(\ref{eq:wf1}) with the momentum shifted from ${\bf k}_0=(\pi,0)$ to $(0,0)$. Note that the above VMC and DMRG calculations are carried out at a finite ladder size $N=40\times2$ under the open boundary condition. }  
\end{figure}

On the other hand, according to the DMRG calculation \cite{DMRG3}, the hopping energy $E_t$ of the single hole shows a ``quantum critical point'' at $\alpha_c\approx0.7$ as indicated by the second derivative over $\alpha$ (see Fig.~\ref{fig:Et} and the inset). By comparison, the corresponding kinetic energy of $|\Psi_\mathrm G\rangle_{\mathrm{1h}} $ is also shown in Fig.~\ref{fig:Et}, which exhibits a singularity at $\alpha_c\approx 0.7$ (see below) indicated by the red arrow, very close to that of the DMRG \cite{DMRG3}.

\begin{figure}[h]
\includegraphics[width=8.5cm,height=6.36cm]{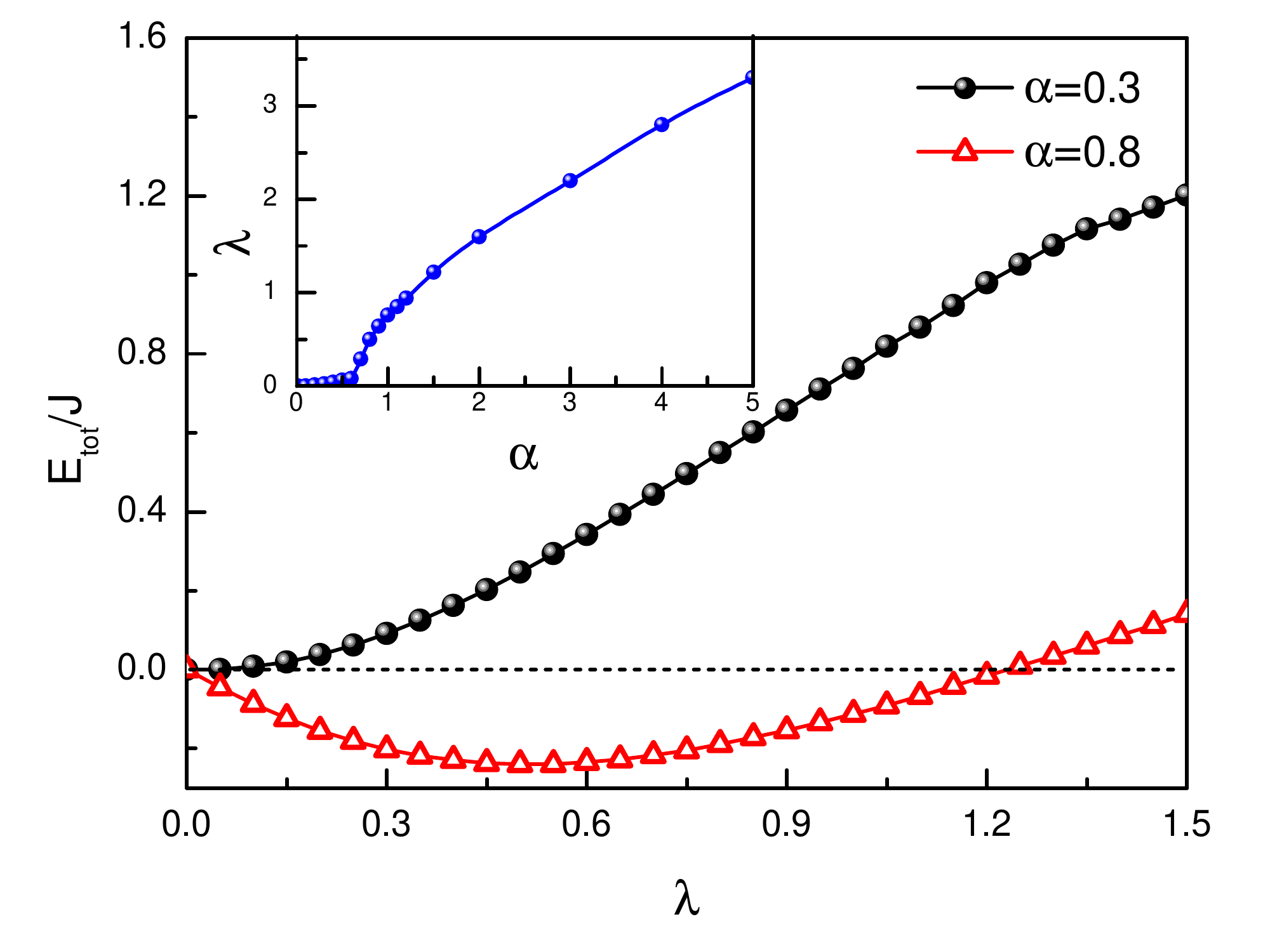}
\caption{\label{fig:2nd_order}(color online). The critical transition at $\alpha_c$ determined by the non-Bloch-like wave function Eq.~(\ref{eq:wf}) as shown in Fig.~\ref{fig:Et}: When $\alpha<\alpha_c$ ($\alpha>\alpha_c$), the minimal total energy $E_\mathrm{tot}$ corresponds to $\lambda\rightarrow 0$ ($\lambda\neq0$). The inset shows the ``order parameter'' $\lambda$ as a function of $\alpha$ [$E_\mathrm{tot}(\lambda=0)$ is rescaled to zero] under a finite ladder size $N=40\times2$. }
\end{figure}

To look more closely, in Fig.~\ref{fig:2nd_order}, $E_\mathrm{tot}$ of $|\Psi_\mathrm G\rangle_{\mathrm{1h}}$ as a function of the variational parameter $\lambda$ is presented at two typical values of $\alpha$ ($\alpha=0.3<\alpha_c$ and $\alpha=0.8>\alpha_c$). One finds that the energy minimum takes place at $\lambda \sim 0$ if $\alpha<\alpha_c$, and $\lambda\neq0$ if $\alpha>\alpha_c$. The inset of Fig.~\ref{fig:2nd_order} further shows $\lambda$ vs. $\alpha$. The systematic change of $E_{\mathrm{tot}}(\lambda)$ with respect to $\alpha$ thus resembles the Ginzburg-Landau theory of second order phase transition, with $\alpha$ playing the role of the ``temperature'' and $\lambda$ the ``order parameter''. Since a finite size calculation is involved here [Fig.~\ref{fig:2nd_order}], one needs to examine more carefully the distinct behaviors on the two sides of $\alpha_c$ in the following.

In contrast to the continuous transition at $\alpha_c$ found in the DMRG and the ground state $|\Psi_\mathrm G\rangle_{\mathrm{1h}} $, the kinetic energy of the Bloch state $|{\bf k}\rangle_{\mathrm{BL}}$ [defined in Eq.~(\ref{eq:wf1})] shows instead an abrupt change (level crossing) from the momentum ${\bf k}_0=(\pi, 0)$ to ${\bf k}_0=(0, 0)$ at $\alpha_c^\mathrm{BL}\approx0.6$ with the increase of $\alpha$, which is presented in Fig.~\ref{fig:Et} by the dashed curves. Such a \emph{first order transition} is simply due to the sign change of the effective hopping parameter $t_\mathrm{BL}^x$ with the decrease of $\langle\mathrm{RVB}|{\bf S}_i\cdot{\bf S}_{i+\hat x}|\mathrm{RVB}\rangle$ in Eq.~(\ref{eq:etBL}).

\begin{figure}[h]
\includegraphics[width=8cm,height=12.2cm]{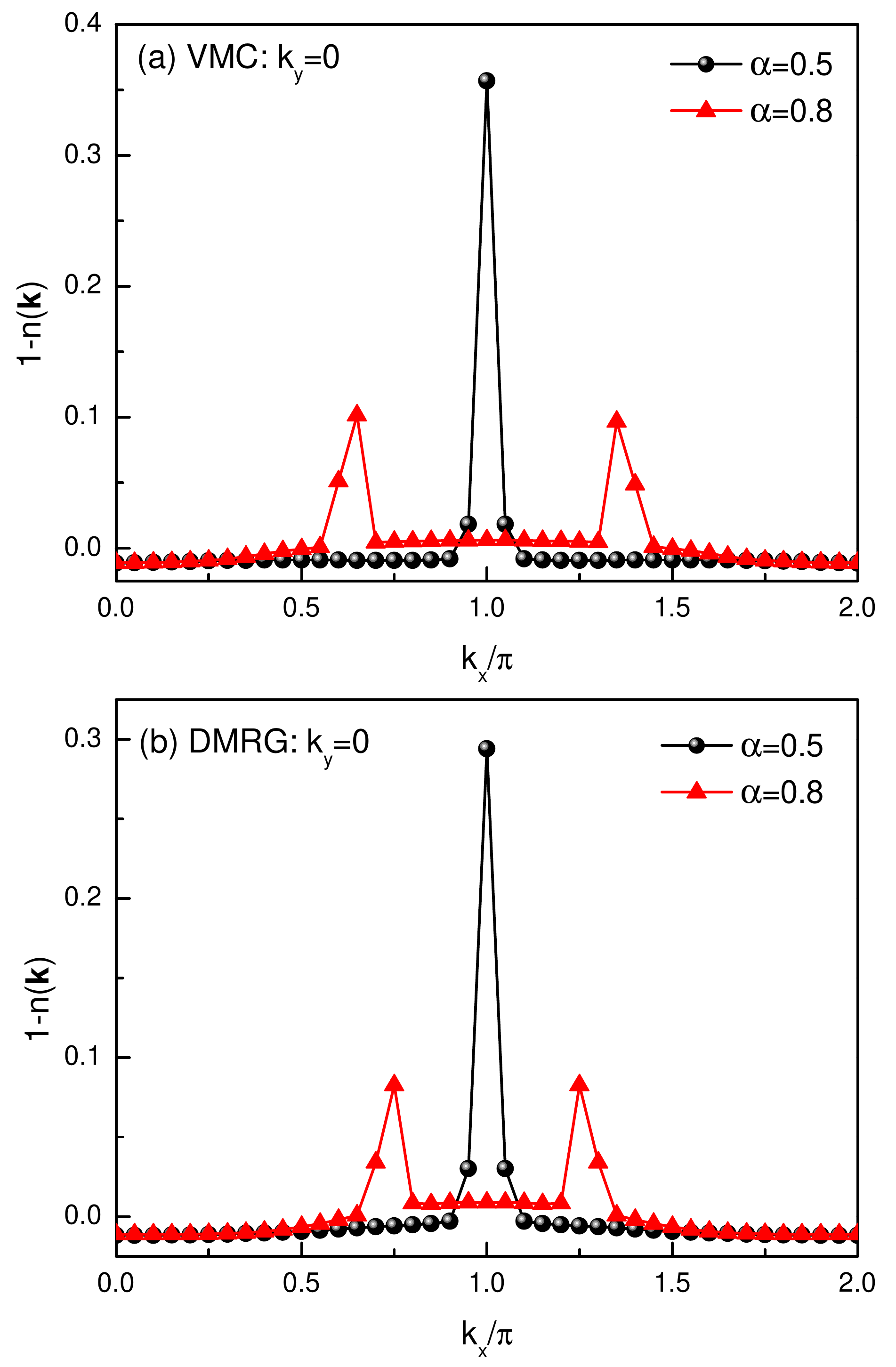}
\caption{\label{fig:nk}(color online). Typical hole momentum distribution $1-n_\bold k$ for $\alpha<\alpha_c$ and $\alpha>\alpha_c$: (a) the VMC calculation; (b) the DMRG calculation. Here each peak in (a) and (b) denotes a characteristic momentum, which is split from the commensurate momentum ${\bf k}_0=(\pi, 0)$ at $\alpha<\alpha_c$ to $k_{x} = {\pi} \pm \kappa$ ($k_y=0$) at $\alpha>\alpha_c$.}
\end{figure}
\begin{figure}[h]
\includegraphics[width=8.5cm,height=6.38cm]{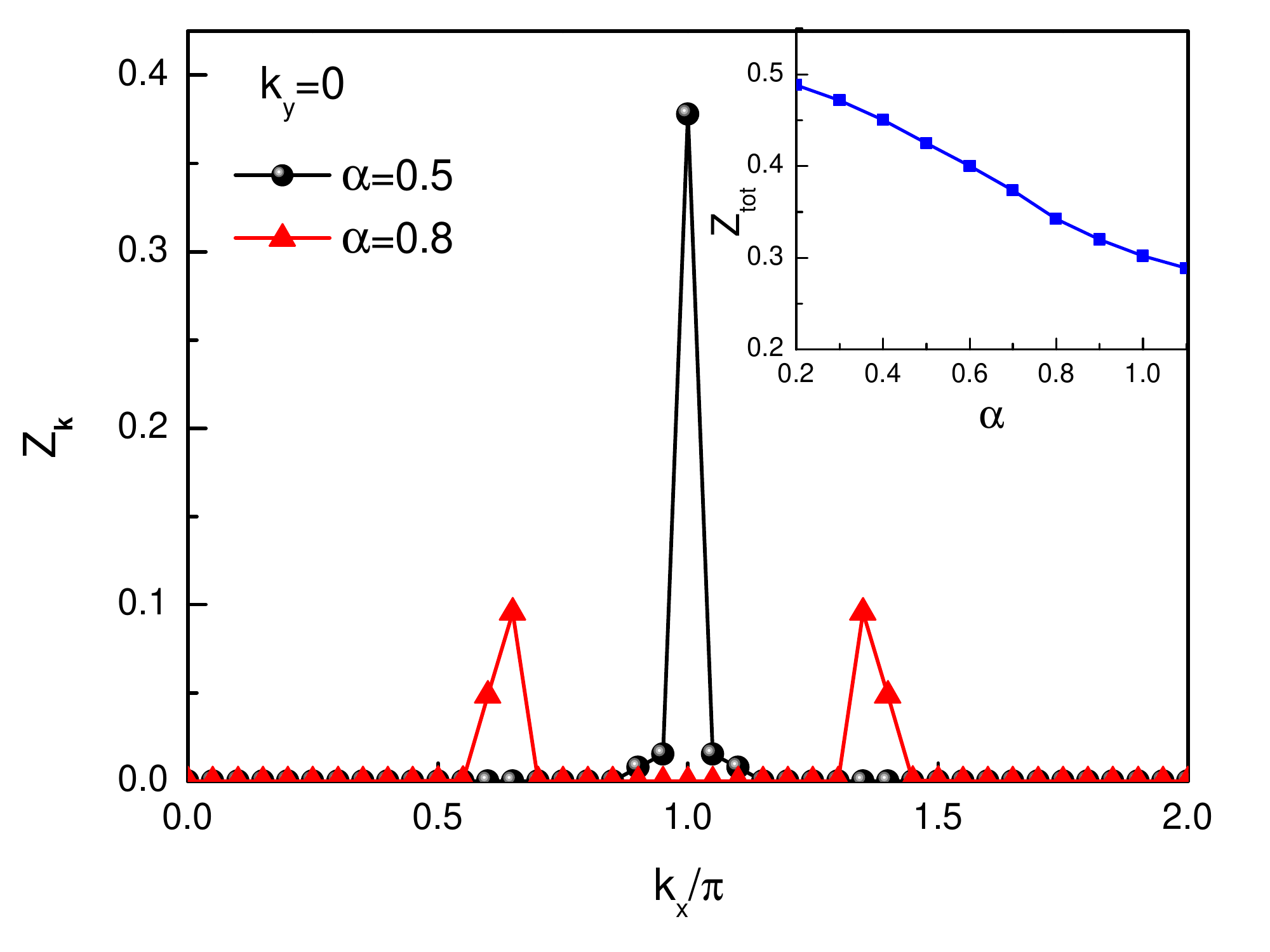}
\caption{\label{fig:Zk}(color online). Quasiparticle weight $Z_\bold k$ at $\alpha=0.5<\alpha_c$ and $\alpha=0.8>\alpha_c$, respectively. The peak position indicates that the characteristic momentum is shifted from the commensurate ${\bf k}_0=(\pi, 0)$ at $\alpha<\alpha_c$ to $k_{x} = {\pi} \pm \kappa$ ($k_y=0$) at $\alpha>\alpha_c$. The inset shows that $Z_\mathrm{tot}\equiv \sum_\mathbf k Z_\mathbf k$ is smooth and nonzero across $\alpha_c$ as a function of $\alpha$ (see the text), which is consistent with the DMRG result \cite{Kivelson}.}
\end{figure}

\subsection{Bloch-wave behavior at $\alpha<\alpha_c$} 

Let us examine the nature of physics on the two sides of $\alpha_c$ in detail. It has been found by DMRG \cite{DMRG3} that the hole momentum distribution $1-n_\mathbf k$ ($n_\mathbf k\equiv\sum_{\sigma}\langle c^{\dagger}_{\mathbf k\sigma}c_{\mathbf k\sigma}\rangle$) is peaked at momentum $\mathbf k_0=(\pi,0)$ at $\alpha<\alpha_c$, which is then split into two peaks at $\alpha>\alpha_c$. Very similar properties are found for $|\Psi_\mathrm G\rangle_{\mathrm{1h}} $ in the VMC calculation [see Fig.~\ref{fig:nk} (a)], which are in good agreement with the DMRG results [Fig.~\ref{fig:nk} (b)].

It indicates that at least at small $\alpha$ ($<\alpha_c$), $|\Psi_\mathrm G\rangle_{\mathrm{1h}} $ and $|{\bf k}_0\rangle_{\mathrm{BL}}$ may describe the same quasiparticle state. Note that the main distinction between the variational wave function Eq.~(\ref{eq:wf}) and the Bloch wave function Eq.~(\ref{eq:wf1}) lies in the many-body phase factor $e^{-i\hat\Omega_i }$ appearing in the former. To compare these two wave functions, we study the wave function overlap defined by
\begin{align}\label{eq:a}
a_\mathrm{\bf k} & \equiv\, _{\text {BL}} \langle \mathrm{\bf k} |\Psi_\mathrm G \rangle_{\mathrm{1h}}\\ \nonumber
&= \sqrt{\frac{2}{N}}\sum_i e^{-i\mathrm{\bf k}\cdot {\bf r}_i}  \varphi_h(i)\langle \mathrm{RVB}| e^{-i\hat\Omega_i } n_{i\downarrow} |\mathrm{RVB} \rangle.
\end{align}
Correspondingly the ``quasiparticle spectral weight" is defined by $Z_\mathrm{\bf k}\equiv |a_\mathrm{\bf k}|^{2}$, which measures the probability of finding the bare hole state $|{\bf k}\rangle_{\mathrm{BL}}$ in the ground state $|\Psi_\mathrm G\rangle_{\mathrm{1h}} $ (see Appendix~\ref{Appen:Zk}). Then the ground state of the variational wave function Eq.~(\ref{eq:wf2}) may be reexpressed as follows
\begin{equation}\label{eq:nonBL}
 |\Psi_\mathrm G \rangle_{\mathrm{1h}} =a_\mathrm{\bf k} |\mathrm{\bf k} \rangle_{\text {BL}} + \cdots,
\end{equation}
where the second term $\cdots$ on the right-hand-side (rhs) refers to the non-Bloch-like part that is orthogonal to the bare hole state in the first term.

We have shown that using the first term alone, i.e., the bare hole state which is a Bloch-like state, in the variational procedure, will result in a commensurate momentum at $\mathbf k_0=(\pi,0)$ at small $\alpha$ ($<\alpha_c$). The two wave functions will thus have a finite overlap $Z_{\mathrm{\bf k}} $ at $\mathbf k =\mathbf k_0$ (cf. Fig.~\ref{fig:Zk}), implying that $|\Psi_\mathrm G \rangle_{\mathrm{1h}} $ has the same momentum ${\bf k}_0$. Indeed, our VMC calculation shows that
\begin{equation}\label{eq:avO}
\langle \mathrm{RVB}| c_{i\downarrow}^\dagger e^{-i\hat\Omega_i } c_{i\downarrow} |\mathrm{RVB} \rangle \propto e^{\pm i \mathrm{\bf k}_0\cdot {\bf r}_i}
\end{equation}
[which may be understood analytically as $\lambda\rightarrow 0$ in Eq.~(\ref{eq:theta})] and $\varphi_h(i) \propto \text{constant} $ to result in $Z_{\mathrm{\bf k}_0}\neq 0$. In this regime, the distinction between $|\Psi_\mathrm G \rangle$ and $|\mathrm{\bf k}_0 \rangle_{\text {BL}}$, i.e., the second term on the rhs of Eq.~(\ref{eq:nonBL}), is mainly responsible for the effective mass and kinetic energy renormalization \emph{without} changing the momentum ${\bf k}_0$.

\begin{figure}[t]
\includegraphics[width=7.9cm,height=11.56cm]{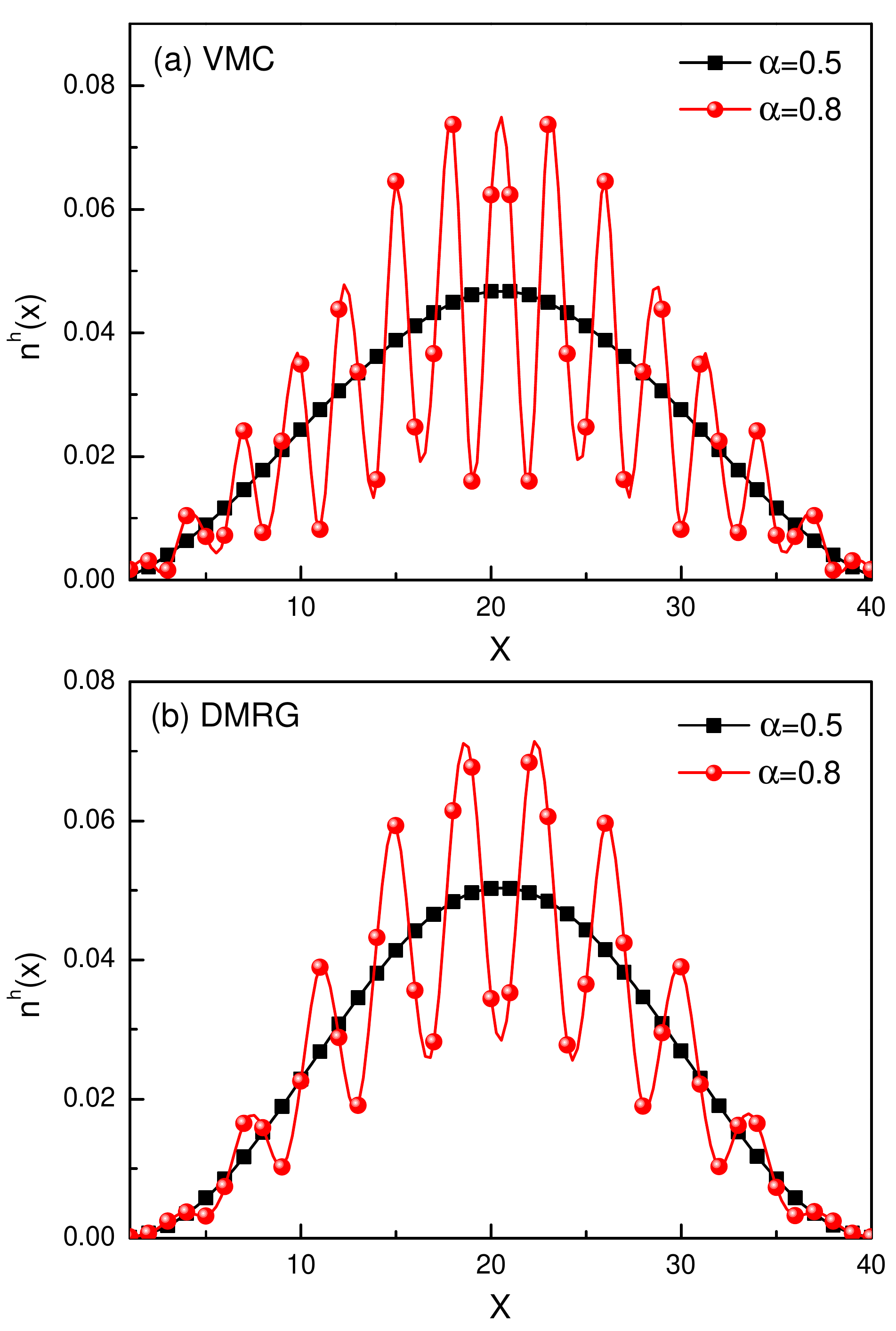}
\caption{\label{fig:nh}(color online). The hole density $n_i^h=|\varphi_h(i)|^2/2$ is smooth for $\alpha<\alpha_c$, and oscillating for $\alpha>\alpha_c$ under an open boundary condition. Note that $n^h_i=n^h_{(x,y)}$ is independent of $y$ and $n^h(x)=\sum_y n^h_{(x,y)}$.}
\end{figure}

\begin{figure}[t]
\includegraphics[width=8cm,height=11.56cm]{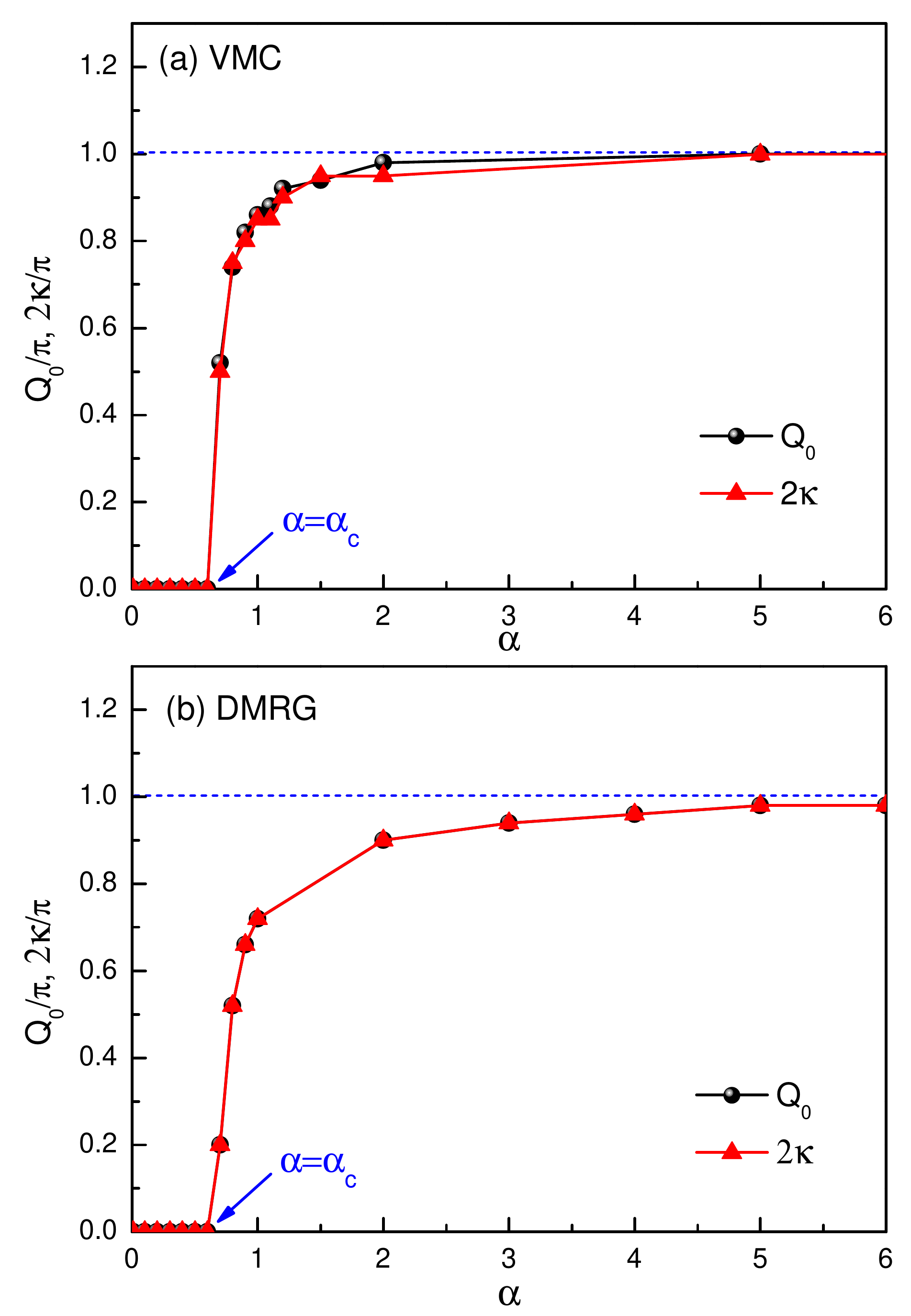}
\caption{\label{fig:kappa,Q0}(color online). Two characteristic momenta $Q_0$ and $2\kappa$ turn out to coincide with each other: $Q_0$ is the charge modulation momentum; $2\kappa$ is defined as the double-peak distance in $n_\bold k$ or $Z_\bold k$. They vanish for $\alpha<\alpha_c$ and approach $\pi$ for $\alpha \gg \alpha_c$.}
\end{figure}

Thus, at $\alpha<\alpha_c$, even though the ground state energy may get further improved by the phase string factor $e^{-i\hat\Omega_i }$ as shown in Fig.~\ref{fig:Et}, the Bloch wave description Eq.~(\ref{eq:wf1}) still remains qualitatively valid with a correct momentum ${\bf k}_0$. It is consistent with the general Landau's paradigm that the quasiparticle wave function has a finite overlap with the Bloch wave function of a bare hole, sharing the same quantum numbers including the one-to-one correspondence of the momentum. As emphasized before, the featureless spin-polaron effect, which is neglected in $|\Psi_\mathrm G \rangle$, may improve further the kinetic energy, but is not expected to change the above one-to-one correspondence of the momentum at ${\bf k}_0$. 

\subsection{Charge modulation as the fingerprint of translational symmetry breaking at $\alpha>\alpha_c$}

The translational symmetry is underlying the above-discussed Bloch-wave-like description of the doped hole at $\alpha<\alpha_c$. But such a symmetry will be found broken at $\alpha>\alpha_c$ in the non-Bloch-like wave function Eq.~(\ref{eq:wf}).

By VMC one finds that the hole density $n^h_i$ is smooth at $\alpha<\alpha_c$, but becomes oscillating at $\alpha>\alpha_c$ as illustrated in Fig.~\ref{fig:nh} (a).
Here the hole density $n_i^h$ can be related to the variational hole wave function $\varphi_h(i)$ as follows (see Appendix~\ref{Appen:nk})
\begin{align}
  \label{}
  n_i^h \equiv 1-\sum_\sigma\langle c_{i\sigma}^\dagger c_{i\sigma} \rangle = \frac{1}{2}\, |\varphi_h(i)|^2.
\end{align}
The DMRG results are shown in Fig.~\ref{fig:nh} (b) for the same parameters, and one finds that the VMC and DMRG are in a qualitative agreement.

The corresponding charge modulation wavevector $Q_0$ is shown in Fig.~\ref{fig:kappa,Q0} (a), which vanishes at $\alpha\rightarrow \alpha_c$ and approaches $\pi$ in the large $\alpha$ limit. Here $Q_0$ well matches with the momentum shift $2\kappa$ between the two peaks in Fig.~\ref{fig:nk}, where the original momentum peak at ${\bf k}={\bf k}_0$ is split into double peaks at ${\bf k}_{\pm} \equiv {\bf k}_0\pm (\kappa,0)$ at $\alpha>\alpha_c$. These results are once again well consistent with the DMRG results\cite{DMRG4} presented in Fig.~\ref{fig:kappa,Q0} (b).

The charge modulation wavevector $Q_0$ can be therefore used to quantify the qualitative change (``phase transition'') of the ground state $|\Psi_\mathrm G \rangle_{\mathrm{1h}} $ at $\alpha_c$, which may be more physical than the variational parameter $\lambda$ shown in Fig.~\ref{fig:2nd_order}. Here the spatial oscillation of the charge density can be traced back to the flux structure of $\tilde{t}_{ij}$ defined in the step-(3) of the variational procedure in Sec.~\ref{sec:Emin} [cf. Eq.~(\ref{eq:et})]. At $\alpha>\alpha_c$, the mean-field solution of $\varphi_h(i)$ becomes oscillating with breaking translational symmetry to imply the unscreened phase string effect (see below).

\subsection{Breakdown of Landau-type quasiparticle description at $\alpha>\alpha_c$}

As already seen previously, the overlap between $ |\Psi_\mathrm G \rangle_{\mathrm{1h}}$ and $|\mathrm{\bf k}_0 \rangle_{\text {BL}} $ disappears, i.e., $Z_{\mathrm{\bf k}_0}=0$ at $\alpha>\alpha_c$ because the momentum in the former is split into incommensurate peaks ${\bf k}_{\pm}$ which no longer coincide with the commensurate $\mathbf k_0$ at $(0,0)$. According to $Z_\mathbf{k}$ shown in Fig.~\ref{fig:Zk}, the momentum is shifted to the incommensurate positions $\mathbf k_\pm$ instead. Here the distinction between $ |\Psi_\mathrm G \rangle_{\mathrm{1h}}$ and a Bloch state with momenta shifted to $\mathbf k_{\pm}$ is crucial. In order to get the correct momentum $\mathbf k_{\pm}$, one cannot simply start with the bare hole or the Bloch state $|\mathbf k\rangle_{\text {BL}} $ in the first term of Eq.~(\ref{eq:nonBL}). As noted above, one would always find a commensurate $\mathbf k_0$ \emph{if} the Bloch state $|\mathbf k\rangle_{\text {BL}} $ is to be used alone variationally or as a self-consistent mean-field state. Rather one has to utilize the \emph{full} form of $ |\Psi_\mathrm G \rangle_{\mathrm{1h}}$ including the non-Bloch-term denoted by $\cdots$ on the rhs of Eq.~(\ref{eq:nonBL}). Therefore it is no longer possible to ``adiabatically connect'' $ |\Psi_\mathrm G \rangle_{\mathrm{1h}}$ with $|\mathbf k_\pm\rangle_{\text {BL}} $ at $\alpha>\alpha_c$, because the latter cannot get the energy (involving nearest-neighbor hopping process) nor momentum (long-wavelength physics) right as a stable mean-field/variational state. This clearly signals the breakdown of Landau's one-to-one correspondence assumption of the momentum for the quasiparticle.

It is instructive to further examine how the Landau-type quasiparticle picture breaks down even though $Z_\mathrm{tot} \neq 0$ as shown in the inset of Fig.~\ref{fig:Zk}, which does not show any singularity at $\alpha=\alpha_c$, consistent with the DMRG \cite{Kivelson}. Here $Z_\mathrm{tot}$ is defined by
$Z_\mathrm{tot}\equiv \sum_\mathbf k  Z_\mathbf k = \sum_i | \langle \mathrm {RVB} |c_{i\downarrow}|\Psi_\mathrm G\rangle |^2 $, which measures the probability of the true ground state remaining in a bare hole state. It can be further expressed as
\begin{equation}\label{eq:ztot}
Z_\mathrm{tot} = \sum_i |\varphi_h(i)|^2|\langle \mathrm {RVB}|c_{i\downarrow}^\dagger e^{-i\hat\Omega_i } c_{i\downarrow} |\mathrm {RVB} \rangle|^2 .
\end{equation}
In evaluating $a_{\mathrm{\bf k}}$ in Eq.~(\ref{eq:a}) or $Z_\mathrm{tot}$ in Eq.~(\ref{eq:ztot}), $e^{-i\hat\Omega_i }$ is averaged over the half-filling state $|\mathrm {RVB}\rangle$, which gives rise to a trivial numerical oscillator similar to Eq.~(\ref{eq:avO}) even at $\alpha>\alpha_c$ to result in a finite $Z_\mathrm{tot} $.  

On the other hand, the incommensurate splitting of ${\bf k}_{\pm}$ is decided by the wave function $\varphi_h(i)$ as the solution of Eq.~(\ref{eq:et}) in the variational procedure. Note that in Eq.~(\ref{eq:et}), the effective hopping integral $\tilde{t}_{ij}$ in Eq.~(\ref{tildet}) may be expressed analytically via Eqs.~(\ref{eq:wf}) and (\ref{eq:wf2}) as $ \tilde{t}_{ij}\equiv \langle\mathrm {RVB}|\hat{t}_{ij} |\mathrm {RVB}\rangle$, with
\begin{equation}\label{eq:hatt}
\hat{t}_{ij}\equiv {t}^0_{ij} \hat{H}_{ij},
\end{equation}
where ${t}^0_{ij}\approx \alpha_{ij}t \left(\frac 1 4-\frac 1 3 \langle\mathrm{RVB}|{\bf S}_i\cdot{\bf S}_j|\mathrm{RVB}\rangle\right)$ is accompanied by a phase factor
\begin{equation}\label{eq:H}
\hat{H}_{ij}\equiv e^{i(A^s_{ij}-\phi^0_{ij})}.
\end{equation}
Here the phase-string factor $e^{-i\hat\Omega_i }$ comes into the crucial play: its phase difference during the nearest neighbor hopping gives rise to a nontrivial flux per plaquette in $\hat{H}_{ij}$ via the gauge link variable \cite{Weng2011a} 
\begin{equation}\label{eq:as}
A^s_{ij}-\phi^0_{ij}\equiv \sum_{l\neq i,j}(\theta_i(l)-\theta_j(l))n_{l\downarrow}
\end{equation}
with $n_{l\downarrow}=c^{\dagger}_{l\downarrow}c_{l\downarrow}$. In the present variational approach, one can numerically determine the flux associated with $\tilde{t}_{ij}$ with the solution $\varphi_h(i)$ exhibiting the charge modulation shown in the last subsection. 

Upon a careful examination, one finds that the effective flux will be sensitive to the spins near the hole, associated with the hole creation in $c_{i\downarrow}|\mathrm{RVB}\rangle$,
as the rest of the spins are in the short-range RVB paired vacuum whose contribution to $e^{-i\hat\Omega_i }$ effectively diminishes away from the hole site. At $\alpha\ll 1$, the RVB pairs are mostly rung-paired such that the spin partner of the doped hole is sitting at the same rung of the hole. In this limit, one has $\lambda\rightarrow 0$ and the effective flux vanishes to result in a translational invariant state. At a larger $\alpha$, the separation between the hole and its spin partner gets enlarged so that $e^{-i\hat\Omega_i }$ becomes nontrivial with $\lambda\neq 0$ (i.e., the phase string becomes unscreened \cite{DMRG3,DMRG4}) to result in the new phase at $\alpha>\alpha_c$. 

Therefore, the single-hole ground state is indeed correctly described by the variational wave function Eq.~(\ref{eq:wf}) rather than the Bloch-like one Eq.~(\ref{eq:wf1}) at $\alpha>\alpha_c$, where the phase-string factor $e^{-i\hat\Omega_i }$ is crucial in regulating the singular short-range hopping process to optimize the ground state energy. In particular, the incommensurate splitting in momentum ${\bf k}_{\pm}$ is decided by $\varphi_h(i)$ with $e^{-i\hat\Omega_i }$ playing an indispensable role in the variational solution. Furthermore, due to the flux effect associated with the effective hopping integral $\tilde{t}_{ij}$, the single-hole's translational symmetry is generally broken. 

\subsection{Beyond the simple variational theory: Localization }
\label{sec:beyond}

\begin{figure*}[t]
\includegraphics[width=18.00cm,height=8.69cm]{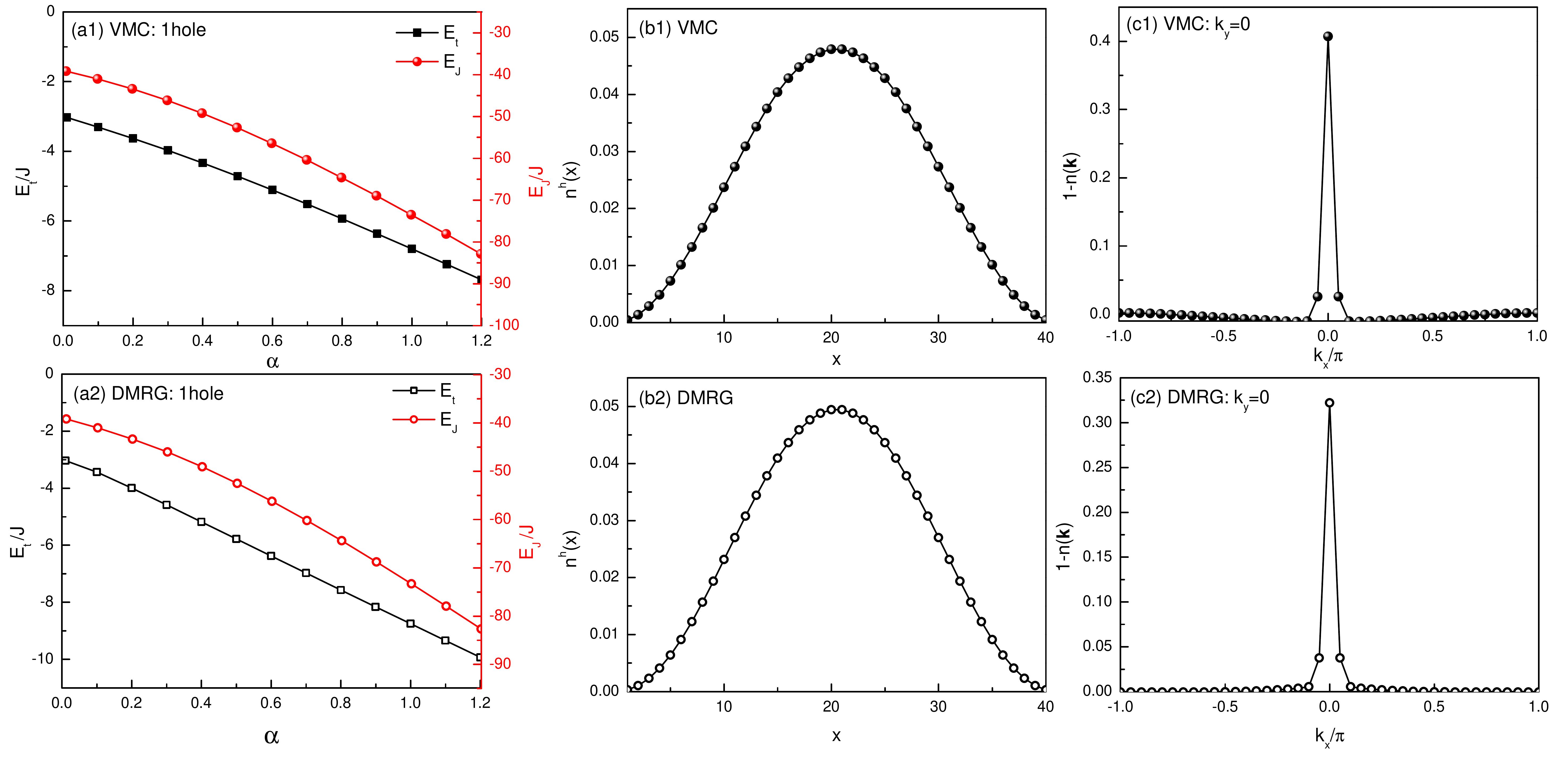}
\caption{\label{fig:stJ}(color online). Absence of the phase string sign structure: VMC and DMRG results for the $\sigma$$\cdot$$t$-$J$ model. (a) Hopping and superexchange energies as a function of $\alpha$. (b) Hole density $n_i^h$ shows no charge modulations at $\alpha=1$. (c) Momentum distribution $1-n_\mathbf k$ has a peak at $\mathbf{k}=(0,0)$ at $\alpha=1$.}
\end{figure*}

At $\alpha>\alpha_c$, with the breakdown of the one-to-one correspondence, the incommensurate momenta ${\bf k}_{\pm}$'s are no longer ``protected'' as in the Landau's quasiparticle description, which may be subject to fluctuations beyond the variational/mean-field theory. To go beyond the variational or mean-field approach, one may generally consider $\varphi_h(i)$ as a \emph{many-body} wave function in Eq.~(\ref{eq:wf}). Then the hopping energy $\langle H_t \rangle$ may be rewritten as $\langle H_t \rangle= \langle\mathrm {RVB}|\hat{H}_h|\mathrm {RVB}\rangle$ with
\begin{equation}\label{eq:eh}
H_h = - \sum_{\langle ij \rangle} \hat{t}_{ij}  \varphi_h^\dagger(j)\varphi_h(i) + \mathrm{h.c.}
\end{equation}
Compared to the previous variational scheme in determining the single-hole wave function $\varphi_h(i)$ in Eq.~(\ref{eq:et}), now the wave function $\varphi_h(i)$ becomes a \emph{many-body} one that is directly subject to the gauge flux [Eq.~(\ref{eq:as})] appearing in $\hat{t}_{ij}$ \emph{before} the average over $|\mathrm {RVB}\rangle$ is taken. The flux enclosed within a plaquette as contributed by Eq.~(\ref{eq:as}) is given by \cite{Weng2011a}
\begin{equation}\label{eq:flux}
\hat{\phi}_{\Box} \equiv \sum_{\Box}(A_{ij}^{s}-\phi^0_{ij})=\pi \sum_{l\in \Box}\left(n_{l\downarrow }-1 \right).
\end{equation}

Besides an average flux $\bar{\phi}_{\Box} $, one may estimate the fluctuation $\delta {\phi}_{\Box} \equiv \hat{\phi}_{\Box} -\bar{\phi}_{\Box} $ by $\langle\mathrm {RVB}|(\delta {\phi}_{\Box})^2|\mathrm {RVB}\rangle$. Since this is a quasi-one-dimensional system, such short-range plaquette flux fluctuations at $\lambda\neq 0$ are generally expected to cause the Anderson-like localization of the doped hole \cite{localization}. In other words, any unscreened scattering between the two characteristic momenta ${\bf k}_{\pm}$ at $\alpha>\alpha_c$, as caused by $\delta {\phi}_{\Box}$, will inevitably lead to the self-localization of the charge. This is consistent with the DMRG calculation \cite{DMRG1,DMRG3} for the two-leg ladder under the periodic boundary condition. There it has been shown \cite{DMRG1,DMRG3} that the energy difference caused by the inserting flux into the ribbon shows an oscillation and exponential decay with the ladder length. Such an insensitivity of the doped charge to the inserting flux indicates that the hole becomes phase incoherence in a sufficient large ladder or self-localized as far as the external U(1) gauge field is concerned. The VMC based on Eqs.~(\ref{eq:eh}) and (\ref{eq:flux}) has indeed confirmed \cite{Sun} the charge localization $\alpha>\alpha_c$ in agreement with the DMRG. The self-localization of the hole and the detailed behavior of $Z_{\mathrm{\bf k}}$ near $ \mathrm{\bf k}_{\pm}$ in the large-ladder-length limit will be further discussed elsewhere.

\section{$\sigma$$\cdot$$t$-$J$ model: Absence of the phase string}
\label{sec:stJ}

So far we have focused on the $t$-$J$ model. The phase string sign structure has been considered to be the most essential factor, which gives rise to the phase string operator $e^{-i\hat\Omega_i }$ in the ground state Eq.~(\ref{eq:wf}). Now we consider the case where such a sign structure can be precisely removed \cite{DMRG1}, which results in a modified local Hamiltonian known as the $\sigma$$\cdot$$t$-$J$ model with a distinct hopping term Eq.~(\ref{eq:sHt}). Then the distinction between the $t$-$J$ and $\sigma$$\cdot$$t$-$J$ models can directly tell the singular role of the phase string effect, which has been clearly demonstrated by DMRG simulations \cite{DMRG1,DMRG2,DMRG3,DMRG4} as mentioned before.

As shown in Appendix~\ref{Appen:sigma}, the sign structure of the $\sigma$$\cdot$$t$-$J$ model can be rigorously identified as the Marshall sign \cite{Marshall1955}. It means that after a Marshall-sign transformation, the $\sigma$$\cdot$$t$-$J$ model (in arbitrary dimension and hole concentration) can be transformed to a model with \emph{trivial sign structure} as defined in Ref.~\onlinecite{Wang2014}. According to the Perron-Frobenius theorem, the ground state of this model has non-negative coefficients in the basis satisfying the Marshall sign rule \cite{Marshall1955}. In particular, the Bloch-like wave function Eq.~(\ref{eq:wf1}) with $\mathbf k_0=(0,0)$ satisfies this sign structure.

It means that the one-hole-doped ground state of the $\sigma$$\cdot$$t$-$J$ ladder should be well described by the Bloch-like wave function Eq.~(\ref{eq:wf1}), in which the translational symmetry is expected to be generally maintained (the featureless spin-polaron effect is still negligible due to the spin gap in the background). Fig.~\ref{fig:stJ} clearly illustrates the overall agreement of the Bloch state $|\mathrm{\bf k}_0 \rangle_{\text {BL}}$ at $\mathrm{\bf k}_0=(0,0)$ with the DMRG result for the one-hole-doped $\sigma$$\cdot$$t$-$J$ ladder.

\section{Conclusion}
\label{sec:conclusion}

A single hole injected into a two-leg spin ladder has manifested a series of novel properties as recently revealed by DMRG simulations \cite{DMRG1,DMRG2,DMRG3,DMRG4,Kivelson}. In this work, we have studied such a system based on a variational wave function $|\Psi_\mathrm G\rangle_{\mathrm{1h}}$ in Eq.~(\ref{eq:wf}) using VMC method. An excellent agreement with the DMRG results have been obtained, which suggests that the trial ground state Eq.~(\ref{eq:wf}) has captured the most essential features of such a doped Mott insulator.

The foremost important message delivered in this work is that the phase string sign structure plays a critical role in a doped Mott insulator.
Indeed, by artificially switching off the phase string sign structure in the $t$-$J$ model to result in the $\sigma$$\cdot$$t$-$J$ model, both DMRG and VMC calculations have shown that the exotic properties exhibited in the former model are totally replaced by a conventional Bloch-wave behavior of the doped hole similar to that in a translationally invariant semiconductor.

In essence, the nontrivial phase string effect implies the translational symmetry breaking in a doped Mott insulator. Both DMRG and VMC have shown that such an effect is responsible for the emergent critical point $\alpha_c$ in the anisotropic two-leg $t$-$J$ ladder of the single hole case. The translationally invariant Bloch state of the doped hole only survives in the strong rung regime of $\alpha<\alpha_c$, where the phase string gets ``screened'' with $\lambda\rightarrow 0$ due to a tight binding of the hole with its spin partner moving in the spin gapped vacuum \cite{DMRG3,DMRG4}. The exotic phenomenon arises at $\alpha>\alpha_c$ where the phase string starts to become ``unscreened'' with $\lambda\neq 0$ as the separation between the hole and its spin partner gets enlarged with increasing $\alpha$ \cite{DMRG3,DMRG4}.

The fingerprint of the unscreened phase string is characterized by the emergent charge density modulation at $\alpha>\alpha_c$. Based on $|\Psi_\mathrm G\rangle_{\mathrm{1h}}$, one finds that the hole density modulation is caused by the quantum interference pattern of the phase string effect as a {\it bulk} property, which cannot be reduced to a conventional standing wave due to two counter-propagating Bloch waves under the {\it open boundary condition} \cite{Kivelson}. Even though the ground state at $\alpha>\alpha_c$ is concomitant with the momentum splitting/Fermi surface (point) reconstruction, the Landau's one-to-one correspondence principle nonetheless breaks down here. Indeed, it is no longer meaningful to try to identify the hole state $|\Psi_G\rangle_{\mathrm{1h}}$ in Eq.~(\ref{eq:wf}) with a conventional quasiparticle since an adiabatic connection in the Landau's paradigm is broken down in such a translational symmetry breaking regime. In particular, it has been pointed out that the self-localization of the charge is inevitable at $\alpha>\alpha_c$ based on the variational form of Eq.~(\ref{eq:wf}).

Finally, the similar symmetry breaking state $|\Psi_\mathrm G\rangle_{\mathrm{1h}}$ is in principle applicable to the $t$-$J$ ladders with more legs. It includes the two-dimensional limit, which is relevant to the high-$T_c$ problem in the cuprates. The generalization of the present VMC for the single hole problem is straightforward, although the DMRG convergence gets more and more difficult with the increase of the leg number. A VMC study along this line is currently underway. Furthermore, the DMRG calculation has shown \cite{DMRG2} a strong binding between two doped holes in the two-leg ladder, implying the ground state Eq.~(\ref{scgs-0}), which can be also studied by VMC in the future. 

\begin{acknowledgements}
We acknowledge stimulating discussions with R. Q. He, H. C. Jiang, D. N . Sheng, C. S. Tian, and J. Zaanen.  Work was supported by the NBRC (973 Program, Nos. 2015CB921000 and 2011CBA00108), by Tsinghua University's ISRP, and in part by Perimeter Institute for Theoretical Physics. Research at Perimeter Institute is supported by the Government of Canada through Industry Canada and by the Province of Ontario through the Ministry of Research and Innovation.
\end{acknowledgements}

\ \newpage

\onecolumngrid
\appendix

\section{VMC for single-hole wave function}
\label{Appen:MC_for_1h_wf}

To provide the necessary notations and make this paper more self-contained, we first present the VMC procedure for the (half-filled) RVB state following Ref.~(\onlinecite{LDA1988}) and (\onlinecite{Sandvik2010}). Whereafter, the VMC formulas for the single-hole wave function $|\Psi_G\rangle$ are derived.

The normalization of RVB state Eq.~(\ref{eq:RVB1}) is given by
\begin{align}\label{eq:normalization1}
  \langle\mathrm{RVB}|\mathrm{RVB}\rangle = \sum_{v,v'}w_{v'}w_v\langle v'|v \rangle.
\end{align}
Since $w_{v'}w_v\langle v|v \rangle$ is positive, we can interpret it as a distribution function. The average value of a physical quantity $\hat O$ is
\begin{align}\label{eq:Omean1}
  \langle \hat O\rangle &= \frac{\langle\mathrm{RVB}|\hat O|\mathrm{RVB}\rangle}{\langle\mathrm{RVB}|\mathrm{RVB}\rangle}
    =  \sum_{v,v'}\frac{w_{v'}w_v\langle v'|v \rangle}{\langle\mathrm{RVB}|\mathrm{RVB}\rangle}\frac{\langle v'|\hat O|v\rangle}{\langle v'|v \rangle}.
\end{align}
The quantity $\langle v'|\hat O|v\rangle /\langle v'|v \rangle$ to be averaged in VMC is usually of order one. For $\hat O=\mathbf S_i\cdot\mathbf S_j$, we have
\begin{align}\label{eq:O_average}
  \frac{\langle v'|\mathbf S_i\cdot\mathbf S_j|v\rangle}{\langle v'|v \rangle} = \delta^{\mathrm{loop}}_{ij} (-1)^{i+j} \cdot\frac{3}{4},
\end{align}
where $\delta^{\mathrm{loop}}_{ij} = 0$ or $1$ indicates whether the two sites $i$ and $j$ belong to the same loop in the transposition-graph of dimer covers $v,v'$.

The most time-consuming part of VMC is the loop tracing in calculating the overlap $\langle v'|v\rangle$. One way to circumvent this problem is to sample the overlap in Monte Carlo by introducing an Ising configuration $\sigma$ (we use $\sigma$ instead of $\{\sigma\}$ for simplicity), besides the two dimer covers $v$ and $v'$ \cite{Sandvik2010}. To combine the VB state and the Ising basis, we introduce the notation
\begin{align}\label{eq:basis1}
  |\sigma\rangle\langle\sigma|v\rangle = \delta_{v,\sigma}|v,\sigma\rangle  = \eta_{v,\sigma}|\sigma\rangle,
\end{align}
where $\delta_{v,\sigma}=|\eta_{v,\sigma}|$ and $\eta_{v,\sigma}=\langle\sigma|v\rangle=0,\pm 1$ is zero or the Marshall sign for the ground state wave function of antiferromagnetic Heisenberg model. Now the VB state $|v\rangle$ and the RVB state Eq.~(\ref{eq:RVB1}) can be expressed as
\begin{align}\label{eq:RVB2}
  |v\rangle &= \sum_{\sigma} \delta_{v,\sigma}|v,\sigma\rangle,\\
  |\mathrm{RVB}\rangle &= \sum_{v}w_v |v\rangle = \sum_{v,\sigma}\delta_{v,\sigma}w_v|v,\sigma\rangle.
\end{align}
The summation is constrained in the space where the dimer cover $v$ and Ising bases $\sigma$ are compatible, i.e., $\delta_{v,\sigma}=1$. By using the fact
\begin{align}\label{eq:overlap1}
  \langle v'|v\rangle &= 2^{N_{v,v'}^{\mathrm{loop}}},\\\label{eq:overlap2}
  \langle v',\sigma'|v,\sigma\rangle &= \delta_{\sigma,\sigma'},
\end{align}
the norm of RVB state is now
\begin{align}\label{eq:normalization2}
  \langle\mathrm{RVB}|\mathrm{RVB}\rangle = \sum_{v,v'}w_{v'}w_v\langle v'|v \rangle
  = \sum_{v,v',\sigma}\delta_{v',\sigma}\delta_{v,\sigma}w_{v'}w_v.
\end{align}
Here, $N_{v,v'}^{\mathrm{loop}}$ is the number of loops in the transposition-graph of dimer covers $v,v'$. Note that the right hand side of Eq.~(\ref{eq:overlap1}) is exactly the number of Ising bases compatible with both the dimer covers $v$ and $v'$. As a result, we can sample the mutual compatible triad $\{v,v',\sigma\}$ configuration space to get the expectation value $\langle\hat O\rangle$ in Eq.~(\ref{eq:Omean1}) without explicitly calculating the overlap $\langle v'|v\rangle$:
\begin{align}\nonumber
  \langle \hat O\rangle &= \sum_{v,v',\sigma} \delta_{v',\sigma} \delta_{v,\sigma}
    \frac{w_{v'}w_v}{\langle\mathrm{RVB}|\mathrm{RVB}\rangle}
    \frac{\langle v'|\hat O|v\rangle}{\langle v'|v \rangle}\\
  &= \frac{\left(\sum_{v,v',\sigma} \delta_{v',\sigma} \delta_{v,\sigma}\right)w_{v'}w_v
    \cdot\langle v'|\hat O|v\rangle/\langle v'|v \rangle}
    {\left(\sum_{v,v',\sigma} \delta_{v',\sigma} \delta_{v,\sigma}\right)w_{v'}w_v}.
\end{align}
The same trick is used in VMC simulations of the single-hole wave function $|\Psi_G\rangle$.

\subsection{Single-hole wave function}
\label{Appen:1h_wf}

We introduce the single-hole ``VB'' states by removing a spin (an up spin without loss of generality) at site $h$ from the half-filled VB states:
\begin{align}\label{}
  |h,v\rangle \equiv b_{h\uparrow}|v\rangle = \sum_{\sigma_h}\delta_{v,\sigma_h}|h,v,\sigma_h\rangle,
\end{align}
where $\sigma_h$ is an Ising basis on the lattice without site $h$, while the dimer cover $v$ covers the whole lattice. And by analogy with Eq.~(\ref{eq:basis1}), we use the notation
\begin{align}\label{eq:basis2}
  |h,\sigma_h\rangle\langle h,\sigma_h|h,v\rangle = \delta_{v,\sigma_h}|h,v,\sigma_h\rangle = \eta_{v,\sigma_h}|h,\sigma_h\rangle.
\end{align}
$\delta_{v,\sigma_h}$ is zero whenever $v$ and $\sigma_h$ is not compatible, i.e., $\sigma_h(i)=\sigma_h(j)$ for some dimer $(i,j)\in v$, or the spin of the site $h'$ originally connecting the hole site $h$ is not a down spin: $\sigma_h(h')\neq\downarrow$. And again, $\eta_{v,\sigma_h}=0,\pm1$ is zero or the Marshall sign.

The variational single-hole wave function is obtained from the RVB state Eq.~(\ref{eq:RVB1}) by removing a spin, accompanied with a unitary transformation $\hat\Lambda$:
\begin{align}\label{eq:one_hole_wf_}
  |\Psi_G\rangle = \hat\Lambda \sum_{h} \varphi_h(h) c_{h\uparrow}|\mathrm{RVB}\rangle = \sum_{h,v,\sigma_h}\delta_{v,\sigma_h}\Lambda(h,\sigma_h)\varphi_h(h)w_v|h,v,\sigma_h\rangle.
\end{align}
Here, the hole wave function $\varphi_h(i)$ is normalized as $\sum_i |\varphi_h(i)|^2=1$. The U(1) phase factor $\Lambda(h,\sigma_h)$ is a function of the hole position $h$ and spin configuration $\sigma_h$, and is defined by
\begin{align}
  \label{eq:LambdaFrac}
  \hat\Lambda |h,v,\sigma_h\rangle = \Lambda(h,\sigma_h) |h,v,\sigma_h\rangle = \prod_{l\neq h} \Lambda(h,l,\sigma_h(l))\ |h,v,\sigma_h\rangle.
\end{align}
The phase factor $\Lambda(h,\sigma_h)$ is fractionalized to $\prod_{l\neq h} \Lambda(h,l,\sigma_h(l))$ in the last step of Eq.~(\ref{eq:LambdaFrac}), which is similar to the fractionalization of amplitude in the Liang-Doucot-Anderson type RVB state \cite{LDA1988}. We have different choices of phase factor $\Lambda(h,l,\sigma)$:

(i) If we choose $\Lambda(h,l,\sigma)=1$, then $\hat\Lambda=1$, and $|\Psi_G\rangle$ and $|\Phi_G\rangle$ are the same. 

(ii) For 2D rotational invariant system, we choose $\Lambda(h,l,\sigma)=e^{i\phi_{hl} \delta_{\sigma\downarrow}}$, where $\phi_{hl}=\mathrm{Im} \ln (z_h-z_l)$ and $z_j=x_j+i y_j$ is the complex coordinate of site $j$. 

(iii) For ladder system and $\hat\Lambda$ defined in the mainbody of the paper, we choose an anisotropic phase factor characterized by $\lambda$: $\Lambda(h,l,\sigma)=e^{i\theta_{hl} \delta_{\sigma\downarrow}}$, where $e^{i\phi_{hl}}$ and $e^{i\theta_{hl}}$ are in the same quadrant and $\tan\theta_{hl}=\lambda\tan\phi_{hl}$.

Similar to Eq.~(\ref{eq:normalization2}), the normalization of the single-hole wave function is given by
\begin{align}\label{eq:normalization3}\nonumber
  \langle\Psi_G|\Psi_G\rangle
  &= \sum_{v,v'} \sum_h \left( \sum_{\sigma_h} \delta_{v,\sigma_h} \delta_{v',\sigma_h} \right) |\varphi_h(h)|^2 w_{v'}w_v \\\nonumber
  &= \sum_{v,v'} 2^{N_{v,v'}^{\mathrm{loop}}-1} \cdot w_{v'}w_v \left( \sum_h |\varphi_h(h)|^2 \right) \\
  &= \left( \sum_{v,v',\sigma^0} \delta_{v,\sigma^0} \delta_{v',\sigma^0} \right) \frac{1}{2} w_{v'}w_v,
\end{align}
due to the inner product of our bases
\begin{align}\label{eq:overlap3}
  \langle h',v'|h,v\rangle &= \delta_{h,h'} 2^{N_{v,v'}^{\mathrm{loop}}-1}, \\\label{overlap4}
  \langle h',v',\sigma_{h'}'|h,v,\sigma_{h}\rangle &= \delta_{h,h'}\delta_{\sigma_h,\sigma_h'}.
\end{align}
These equations bear a resemblance to Eq.~(\ref{eq:overlap1}) and Eq.~(\ref{eq:overlap2}). Note that there is a $-1$ in the exponent on the right hand side of Eq.~(\ref{eq:overlap3}), for the transposition-graph loop containing the hole contributes only one Ising configuration rather than two. The number $2^{N_{v,v'}^{\mathrm{loop}}-1}$ is also exactly the number of Ising bases compatible with hole position $h$ and dimer covers  $v,v'$. In the last line of Eq.~(\ref{eq:normalization3}), we put an additional summation over the Ising bases $\sigma^0$ on the whole lattice (without hole), such that the configuration space $\{ (v,v',\sigma^0) \}$ is the same as the half filled case in Eq.~(\ref{eq:normalization2}). Comparing Eq.~(\ref{eq:normalization3}) to Eq.~(\ref{eq:normalization2}), we find that the normalization of $|\Psi_G\rangle$ and $|\mathrm{RVB}\rangle$ are related by
\begin{align}
  \label{eq:normrel}
  \langle\Psi_G|\Psi_G\rangle = \frac{1}{2} \langle\mathrm{RVB}|\mathrm{RVB}\rangle.
\end{align}

\subsection{Superexchange energy}
\label{Appen:MC_HJ}

We start with the expectation value of the Heisenberg superexchange terms, which are easier in the sense that they do not change the hole position. The average superexchange energy between sites $i$ and $j$ is
\begin{align}\nonumber
  \langle H_{ij}^J \rangle &=
    \sum_{v,v'} \frac{w_{v'}w_v} {\langle\Psi_G|\Psi_G\rangle} \sum_h \sum_{\sigma_h,\sigma_h'} \delta_{v,\sigma_h} \delta_{v',\sigma_h'} \cdot|\varphi_h(h)|^2
    \cdot \mathrm{Re}\left(\Lambda^\ast(h,\sigma_h') \Lambda(h,\sigma_h)\right) \cdot \langle h,v',\sigma_h'| H_{ij}^J |h,v,\sigma_h \rangle\\
  &= \frac{ \left( \sum_{v,v',\sigma^0} \delta_{v,\sigma^0} \delta_{v',\sigma^0} \right)
   \frac{1}{2} w_{v'}w_v\cdot E_{ij}^J(v,v')}
    {\left( \sum_{v,v',\sigma^0} \delta_{v,\sigma^0} \delta_{v',\sigma^0} \right)
    \frac{1}{2} w_{v'}w_v},
\end{align}
where
\begin{align}
  \label{eq:E_Jij}
  E_{ij}^J(v,v') &= \sum_h |\varphi_h(h)|^2 E_{ij}^J(h,v,v'), \\\label{eq:E_J_hvv'}
  E_{ij}^J(h,v,v') &= \sum_{\sigma_h,\sigma_h'} \delta_{v,\sigma_h} \delta_{v',\sigma_h'}
  \cdot \mathrm{Re} \left( \Lambda^\ast(h,\sigma_h') \Lambda(h,\sigma_h) \right)
  \cdot   \frac{ \langle h,v',\sigma_h'| H_{ij}^J |h,v,\sigma_h \rangle } {\langle h,v'| h,v \rangle}.
\end{align}
We can now interpret $(1/2)w_{v'}w_v/\langle\Psi_G|\Psi_G\rangle$ as a probability function in the space of compatible configurations $(v,v',\sigma^0)$. The superexchange energy can be calculated by averaging $E_{ij}^J(v,v')$ in standard Monte Carlo procedure. For a fixed configuration, the energy to be averaged $E_{ij}^J(v,v')$ has summations over hole position $h$ and spin configurations $\sigma_h,\sigma_h'$.

The Eq.~(\ref{eq:E_J_hvv'}) can be further simplified. The element $\langle h,v',\sigma_h'| H_{ij}^J |h,v,\sigma_h \rangle$ makes the summations over spin configurations easier as it forces the spin configurations $\sigma_h$ and $\sigma_h'$ are almost the same except on sites $i$ and $j$. Now fix the configuration $(v,v',\sigma^0)$ and the hole position $h$, we should distinguish three different situations to simplify $E_{ij}^J(h,v,v')$:

(i) Hole site $h$ coincides with sites $i$ or $j$, then $E_{ij}^J(h,v,v')=0$.

(ii) Sites $i$ and $j$ belong to different loops in the transposition-graph of dimer covers $v,v'$. For terms $S_i^+ S_j^-$ and $S_i^- S_j^+$ in $H_{ij}^J$, the expectation values are always zero because of the compatibility of the dimer covers and spin configurations (one closed loop can not have a single antiferromagnetic domain wall). For diagonal term $S_i^z S_j^z$, however, although the expectation over a fixed spin configuration is not zero, the summation of these terms is zero due to the independence of $\sigma_h(i)$ and $\sigma_h(j)$.

(iii) Sites $i,j$ belong to the same loop in the transposition-graph. If this loop does not contain the hole site, one can show Eq.~(\ref{eq:E_J_hvv'}) becomes
\begin{align}\nonumber\label{eq:EJij}
  E_{ij}^J(h,v,v') &= 2 \times 2^{N_{v,v'}^{\mathrm{loop}}-2} \cdot \mathrm{Re}(\Delta\Lambda^J_{ij}) \cdot \frac{-J/2}{2^{N_{v,v'}^{\mathrm{loop}}-1}} + 2^{N_{v,v'}^{\mathrm{loop}}-1} \cdot 1 \cdot \frac{-J/4}{2^{N_{v,v'}^{\mathrm{loop}}-1}} \\
  &= -\frac{J}{2} \mathrm{Re} (\Delta\Lambda^J_{ij}(h)) -\frac{J}{4},
\end{align}
where the first term in the first line comes from the equal contributions of $S_i^+ S_j^-$ and $S_i^- S_j^+$, and the second term comes from $S_i^z S_j^z$. The result Eq.~(\ref{eq:EJij}) is still valid, when the sites $i,j,h$ all belong to the same loop (but these three sites are different). But the origins of each term are different: only $S_i^+ S_j^-$ \emph{or} $S_i^- S_j^+$ contributes to the first term.

Now we turn to the definition of phase factor change $\Delta\Lambda^J_{ij}(h)$ in Eq.~(\ref{eq:EJij}), which comes from the phase difference between the bra and ket of the single-hole wave function:

(i) When we choose $\Lambda(h,l,\sigma)=1$ by ignoring the phase factor $\hat\Lambda$ in the single-hole wave function, the phase factor change $\Delta\Lambda^J_{ij}(h)=1$ and Eq.~(\ref{eq:EJij}) becomes $E_{ij}^J(h,v,v')=-3J/4$ when $i,j$ do not coincide with $h$ and belong to the same loop in the transposition-graph, which recovers the result of the half-filled energy expectation value Eq.~(\ref{eq:O_average}).

(ii) For a generic fractionalized phase factor $\Lambda(h,\sigma_h) = \prod_{l\neq h} \Lambda(h,l,\sigma_h(l))$, the phase factor change is
\begin{align}\label{eq:d_phase_J2}
  \Delta\Lambda^J_{ij}(h) = \Lambda^\ast(h,i,\uparrow) \Lambda^\ast(h,j,\downarrow) \Lambda(h,i,\downarrow) \Lambda(h,j,\uparrow).
\end{align}
Rotational symmetry ($\Lambda(h,l,\sigma)=e^{i\phi_{hl} \delta_{\sigma\downarrow}}$) simplifies the above result to
\begin{align}\label{eq:d_phase_J2_rot}
  \left. \Delta\Lambda^{J}_{ij}(h) \right|_{\mathrm{2D}} = e^{i (\phi_{hi}-\phi_{hj})} = \exp \left( i\; \mathrm{Im} \ln \left( \frac{z_h-z_i}{z_h-z_j} \right) \right).
\end{align}

(iii) For ladder system with $\Lambda(h,l,\sigma)=e^{i\theta_{hl} \delta_{\sigma\downarrow}}$ and $\tan\theta_{hl}=\lambda\tan\phi_{hl}$, we can use Eq.~(\ref{eq:d_phase_J2}) and have
\begin{align}
  \label{}
  \left. \Delta\Lambda^{J}_{ij}(h)\right|_{\mathrm{ladder}} = e^{i (\theta_{hi}-\theta_{hj})}.
\end{align}

In summary, the total superexchange energy is evaluated in Monte Carlo by calculating Eqs.~(\ref{eq:E_Jij}), where $E^J_{ij}(h,v,v')$ is zero in some conditions or given by Eq.~(\ref{eq:EJij}) otherwise. Same as the (half-filled) Heisenberg model, the Monte Carlo configuration space is spanned by two VB states and one spin configuration $(v,v',\sigma^0)$ which are compatible, with non-negative weight $w_{v'}w_v/2$.

\subsection{Hopping energy}
\label{Appen:MC_Ht}

Now turn to the expectation value of the hopping term, which moves the hole from one site to another. Direct calculation shows:
\begin{align}\nonumber
  \langle H_t \rangle
  &=
    \sum_{v,v'} \frac{w_{v'}w_v} {\langle\Psi_G|\Psi_G\rangle} \sum_{h,h'} \sum_{\sigma_h,\sigma'_{h'}} \delta_{v,\sigma_h} \delta_{v',\sigma'_{h'}}
    \cdot\mathrm{Re} \left( \varphi_h^\ast (h') \varphi_h(h) \Lambda^\ast(h',\sigma_{h'}') \Lambda(h,\sigma_h) \right) \\\nonumber
  &\quad 
    \cdot \langle h',v',\sigma_{h'}'| H_t |h,v,\sigma_h \rangle \\
  &=
    \frac{\left(\sum_{v,v',\sigma^0}\delta_{v,\sigma_h^0} \delta_{v',\sigma_h^0}\right)
    \frac{1}{2} w_{v'}w_v\cdot E_t(v,v')}
    {\left(\sum_{v,v',\sigma^0}\delta_{v,\sigma^0} \delta_{v',\sigma^0}\right) \frac{1}{2} w_{v'}w_v},
\end{align}
where the averaged quantity in VMC is
\begin{align}
  \label{eq:Etvv'}
  E_t(v,v') &= \sum_{h,h'} E_t(h,h',v,v'), \\\label{eq:Etdown_}
  E_t(h,h',v,v') &= \sum_{\sigma_h,\sigma'_{h'}} \delta_{v,\sigma_h} \delta_{v',\sigma'_{h'}} \cdot \mathrm{Re} ( \varphi_h^\ast (h') \varphi_h(h) \Lambda^\ast(h',\sigma'_{h'}) \Lambda(h,\sigma_h) )
  \cdot  \frac{ \langle h',v',\sigma'_{h'}| H_t |h,v,\sigma_h \rangle } {\langle h,v'| h,v \rangle}.
\end{align}

Now we would like to simplify Eq.~(\ref{eq:Etdown_}). The summation over hole positions $h$ and $h'$ in Eq.~(\ref{eq:Etvv'}) has a constraint that $h$ and $h'$ must be neighbouring sites, otherwise the hopping energy vanishes.
Now fix a Monte Carlo configuration $(v,v',\sigma^0)$ and sites $h,h'$, there are two different situations according to whether sites $h$ and $h'$ belong to the same loop in the transposition-graph of dimer covers $v$ and $v'$:

\begin{figure}[h]
\includegraphics[width=7cm,height=6.5cm]{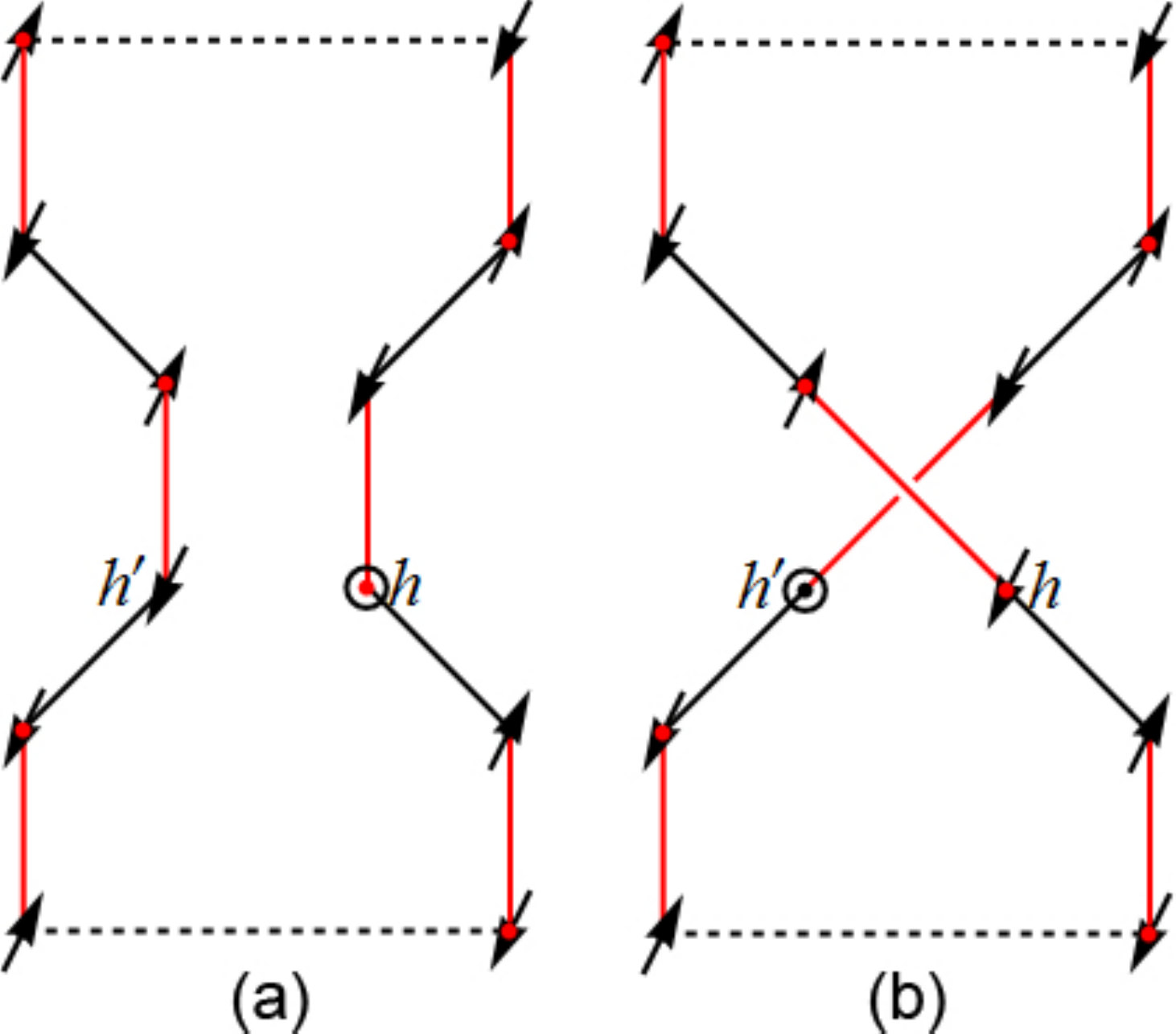}
\caption{\label{fig:Ht_1loop}(color online). Topological graph of hole down-spin exchange process (sites $h$ and $h'$ belong to the same loop). Black and red dots represent sites of different sublattices. Dashed line means the length of the loop is arbitrary. (a) Configuration before hopping. Dimers belong to $v'$ ($v$) are illustrated by black (red) bonds. (b) Configuration after hopping. Spins on this transposition-graph must be alternating in the new loop, with hole understood as an up spin.}
\end{figure}

(i) $h,h'$ belong to the same loop $L_{h,h'}$. The only possible incident under the action of $H_t$ is the exchange of a hole and a down spin (see Fig.~\ref{fig:Ht_1loop}). The orthogonal property Eq.~(\ref{overlap4}) has several constraints: the bond $\langle ij\rangle$ must be the same as $\langle hh'\rangle$; the spin configurations $\sigma_h$ and $\sigma_{h'}'$ must satisfy relations $\sigma_h(h')=\sigma_{h'}'(h)=\downarrow$, $\sigma_h(l)=\sigma_{h'}'(l)$, for $l\neq h,h'$; spin configuration on sites belong to $L_{h,h'}$ is uniquely determined by $(h,h',v,v')$, while for every other loop $L\neq L_{h,h'}$, there are two possible spin configurations. For any given pair of initial and final states with nonzero hopping energy contribution, we have $\langle h',v',\sigma_{h'}'| H^t |h,v,\sigma_h \rangle = -\alpha_{ij} t$ (the Marshall sign difference of the initial and final state cancels the fermion permutation sign). Therefore, the total hopping energy Eq.~(\ref{eq:Etdown_}) is given by
\begin{align}\label{eq:Etdown}
  E_{h\downarrow}^t(h,h',v,v') = -\alpha_{ij} t \cdot \mathrm{Re} \left(\varphi_h^\ast (h') \varphi_h(h) \Delta\Lambda_{h\downarrow}^t(h,h',v,v')\right),
\end{align}
where the total phase factor change $\Delta\Lambda_{h\downarrow}^t(h,h',v,v')$ is the only nontrivial value to be calculated. Follow Eq.~(\ref{eq:LambdaFrac}) and the spin configuration constraints, the fractionalized phase factors can be divide into three parts: $l=h\ \mathrm{or}\ h'$, $l\in L_{h,h'}$, $l\in L\neq L_{h,h'}$. Accordingly, the total phase factor change $\Delta\Lambda_{h\downarrow}^t(h,h',v,v')$ in the hopping process is a product of three phase factors:
\begin{align}\label{eq:phasefactor_h_down}\nonumber
  \Delta\Lambda_{h\downarrow}^t(h,h',v,v')
    &= \Lambda^\ast(h',h,\downarrow) \Lambda(h,h',\downarrow)\cdot
    \left(\prod_{l\in L_{h,h'},l\neq h,h'}
    \Lambda^\ast(h',l,\sigma_h(l)) \Lambda(h,l,\sigma_h(l))\right)\\
    &\cdot\left(\prod_{L\neq L_{h,h'}}
    \frac{1}{2}
    \sum_{\sigma_h(L)=\pm 1}
    \prod_{l\in L}
    \Lambda^\ast(h',l,\sigma_h(l)) \Lambda(h,l,\sigma_h(l))\right).
\end{align}
The spin configuration $\sigma_h(l)$ for $l\in L_{h,h'}$ in the second phase factor is determined by $(h,h',v,v')$. While the summation $\sum_{\sigma_h(L)=\pm 1}$ in the third phase factor sums over two possible spin configurations in a loop $L$ different from $L_{h,h'}$.
Note that this summation totally gives us $2^{N_{v,v'}^{\mathrm{loop}}-1}$ terms as the value of $\langle h,v'|h,v\rangle$ (cf. Eq.~(\ref{eq:overlap3})), which becomes the factor $1/2$ when we move it through the product operator of different loops in the third phase factor.
If we choose $\Lambda(h,l,\sigma)=e^{i\theta_{hl} \delta_{\sigma\downarrow}}$ for ladder system, the phase factor change becomes
\begin{align}
  \label{eq:phasefactor_h_down_rotation}
  \left. \Delta\Lambda_{h\downarrow}^t(h,h',v,v') \right|_\mathrm{ladder} =
    -\left(\prod_{\scriptstyle l\in L_{h,h'}\atop\scriptstyle l\neq h,h'}
    e^{i(\theta_{hl}-\theta_{h'l})\delta_{\sigma_h(l),\downarrow}}
    \right)
    \cdot\left(\prod_{L\neq L_{h,h'}}
    \frac{1}{2}
    \left(\prod_{\scriptstyle l\in L\atop\scriptstyle l\in A}
    +\prod_{\scriptstyle l\in L\atop\scriptstyle l\in B}\right)
    e^{i(\theta_{hl}-\theta_{h'l})}
    \right),
\end{align}
where the minus sign comes from the first phase factor $\Lambda^\ast(h',h,\downarrow) \Lambda(h,h',\downarrow)$, and $A$ and $B$ denote two sublattices.

\begin{figure}[h]
\includegraphics[width=6cm,height=7.6cm]{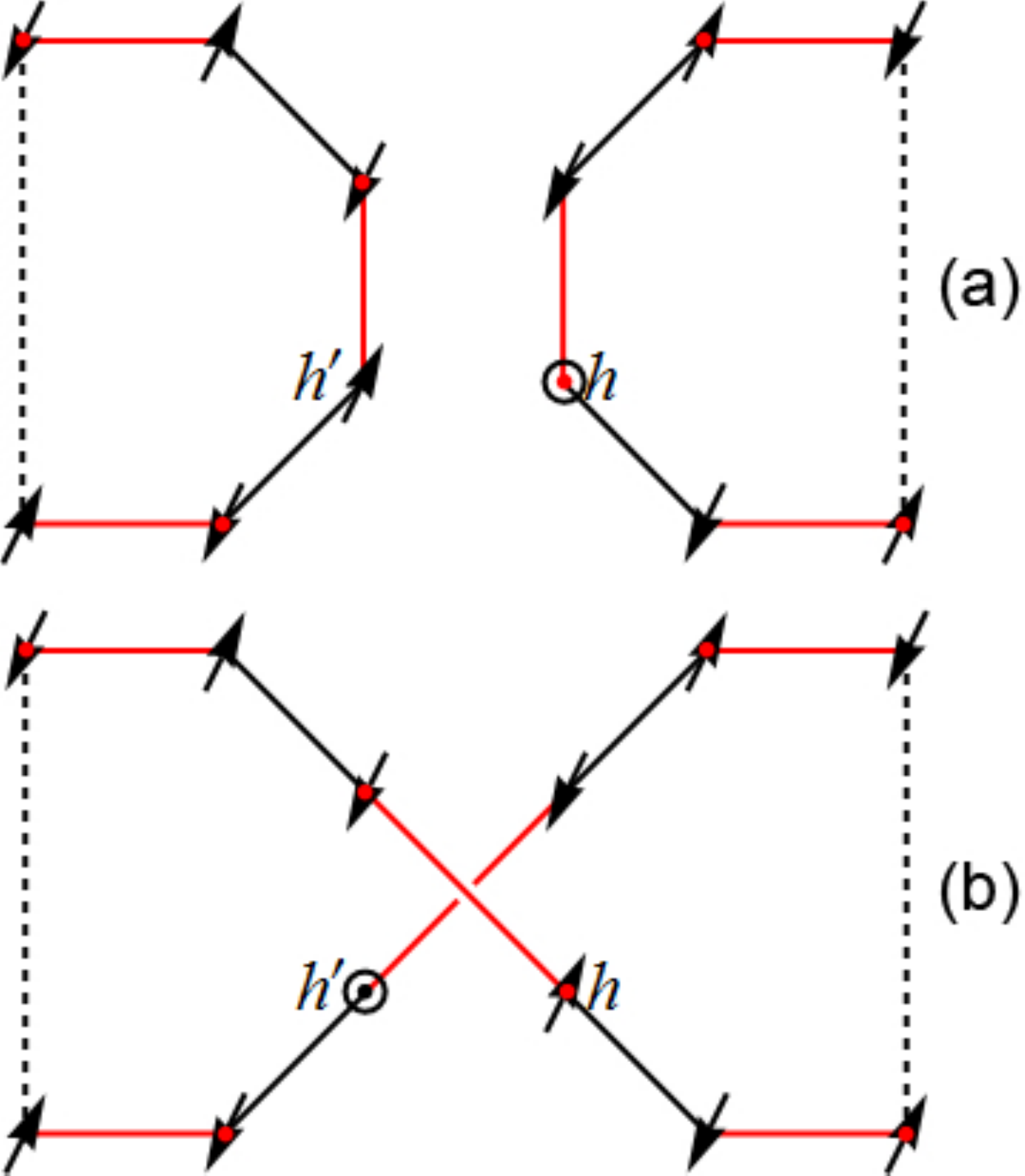}
\caption{\label{fig:Ht_2loop}(color online). Topological graph of hole up-spin exchange process (sites $h$ and $h'$ belong to different loops). (a) Configuration before hopping. (b) Configuration after hopping. Spins on this transposition-graph also must be alternating in the new loop, with hole understood as an up spin.}
\end{figure}

(ii) $h,h'$ belong to different loops $L_h$ and $L_{h'}$. In this case the hole can only exchange with an up spin (see Fig.~\ref{fig:Ht_2loop}). The final results are parallel to the first case: replace spin down by spin up; replace loop $L_{h,h'}$ by loop $L_h$ or $L_{h'}$. The total hopping energy is
\begin{align}\label{eq:Etup}
  E_{h\uparrow}^t(h,h',v,v') = \frac{\alpha_{ij}t}{2}\cdot \mathrm{Re} \left( \varphi_h^\ast (h') \varphi_h(h) \Delta \Lambda_{h\uparrow}^t(h,h',v,v') \right).
\end{align}
Note that there is fermion permutation sign but no Marshall sign difference in the up-spin hopping process. Therefore, there is a minus sign in front of Eq.~(\ref{eq:Etup}) comparing to Eq.~(\ref{eq:Etdown}).
The phase factor change for general $\Lambda(h,l,\sigma)$ and for the ladder system are
\begin{align}\label{phasefactor_h_up}\nonumber
  \Delta\Lambda_{h\uparrow}^t(h,h',v,v')
    &= \Lambda^\ast(h',h,\uparrow) \Lambda(h,h',\uparrow)\cdot
    \left(\prod_{l\in L_h\mathrm{\;or\;}L_{h'},l\neq h,h'}
    \Lambda^\ast(h',l,\sigma_h(l)) \Lambda(h,l,\sigma_h(l))\right)\\
    &\cdot\left(\prod_{L\neq L_h,L_{h'}}
    \frac{1}{2}
    \sum_{\sigma_h(L)=\pm 1}
    \prod_{l\in L}
    \Lambda^\ast(h',l,\sigma_h(l)) \Lambda(h,l,\sigma_h(l))\right),
\end{align}
\begin{align}
  \label{eq:phasefactor_h_up_rotation}
  \left. \Delta \Lambda_{h\uparrow}^t(h,h',v,v') \right|_\mathrm{ladder} =
    \left(\prod_{\scriptstyle l\in L_h\mathrm{\;or\;}L_{h'}\atop\scriptstyle l\neq h,h'}
    e^{i(\theta_{hl}-\theta_{h'l})\delta_{\sigma_h(l),\downarrow}}
    \right)
    \cdot\left(\prod_{L\neq L_h,L_{h'}}
    \frac{1}{2}
    \left(\prod_{\scriptstyle l\in L\atop\scriptstyle l\in A}
    +\prod_{\scriptstyle l\in L\atop\scriptstyle l\in B}\right)
    e^{i(\theta_{hl}-\theta_{h'l})}
    \right),
\end{align}
The minus sign in front of the phase factor change Eq.~(\ref{eq:phasefactor_h_down_rotation}) disappears in Eq.~(\ref{eq:phasefactor_h_up_rotation}) because $\Lambda^\ast(h',h,\uparrow) \Lambda(h,h',\uparrow) = -\Lambda^\ast(h',h,\downarrow) \Lambda(h,h',\downarrow) = 1$ for ladder system. The additional numerical factor $1/2$ in the hopping energy Eq.~(\ref{eq:Etup}) comes from the fact that the overlap of dimer states is $\langle h,v'|h,v\rangle = 2^{N_{v,v'}^{\mathrm{loop}}-1}$, while there are only $N_{v,v'}^{\mathrm{loop}}-2$ loops (those different from $L_h$ and $L_{h'}$) whose spin configurations are not determined.

In summary, the total hopping energy is evaluated in Monte Carlo by calculating Eqs.~(\ref{eq:Etvv'}), (\ref{eq:Etdown}) and (\ref{eq:Etup}). The Monte Carlo configuration space is also spanned by compatible triad $(v,v',\sigma^0)$, which is the same as the configuration space for (half-filled) pure spin model.

\subsection{Momentum distribution}
\label{Appen:nk}

To calculate the momentum distribution $n_\mathbf k$ for $|\Psi_G\rangle$, we should consider long range hopping process:
\begin{align}
  \label{}
  \langle \hat n_{\mathbf k\sigma} \rangle = \langle c_{\mathbf k\sigma}^\dagger c_{\mathbf k\sigma} \rangle =\frac{1}{N} \sum_{i,j} e^{i\mathbf k \cdot (R_i-R_j)} \langle c_{i\sigma}^\dagger c_{j\sigma} \rangle.
\end{align}
We denote the average value of long range hopping process by $T_{ij\sigma}=\langle c_{i\sigma}^\dagger c_{j\sigma} \rangle$. Different from the calculation of the hopping energy which involves only neighbouring sites, $i$ and $j$ in $T_{ij\sigma}$ can be the same site or separate far from each other.
Direct calculation shows:
\begin{align}\nonumber\label{eq:Tijs}
  T_{ij\sigma} = \langle c_{i\sigma}^\dagger c_{j\sigma} \rangle
  &=
    \sum_{v,v'} \frac{w_{v'}w_v} {\langle\Psi_G|\Psi_G\rangle} \sum_{h,h'} \sum_{\sigma_h,\sigma'_{h'}} \delta_{v,\sigma_h} \delta_{v',\sigma'_{h'}}
    \cdot\mathrm{Re} \left( \varphi_h^\ast (h') \varphi_h(h) \Lambda^\ast(h',\sigma_{h'}') \Lambda(h,\sigma_h) \right) \\\nonumber
  &\quad 
    \cdot \langle h',v',\sigma_{h'}'| c_{i\sigma}^\dagger c_{j\sigma} |h,v,\sigma_h \rangle \\
  &=
    \frac{\left(\sum_{v,v',\sigma^0}\delta_{v,\sigma_h^0} \delta_{v',\sigma_h^0}\right)
    \frac{1}{2} w_{v'}w_v\cdot T_{ij\sigma}(v,v')}
    {\left(\sum_{v,v',\sigma^0}\delta_{v,\sigma^0} \delta_{v',\sigma^0}\right) \frac{1}{2} w_{v'}w_v},
\end{align}
where $T_{ij\sigma}(v,v')$ is to be calculated in every Monte Carlo measurement step.

The expressions of $T_{ij\sigma}(v,v')$ are different for $i=j$ and $i\neq j$. Let us consider $i=j$ first:
\begin{align}
  \label{}\nonumber
  T_{ii\sigma}(v,v') &= \sum_{h\neq i} \frac{1}{2^{N_{v,v'}^\mathrm{loop}}} \sum_{\sigma_h} \delta_{v,\sigma_h} \delta_{v',\sigma_h} |\varphi_h(h)|^2 \delta_{\sigma_h(i),\sigma}\\
  &= \sum_{h\neq i} |\varphi_h(h)|^2 \left[ \delta_{h,i}^\mathrm{loop} \left(\delta_{\sigma,\uparrow} \delta_{h,i}^\mathrm{sublatt} + \delta_{\sigma,\downarrow} (1-\delta_{h,i}^\mathrm{sublatt}) \right) + (1-\delta_{h,i}^\mathrm{loop}) \frac{1}{2} \right].
\end{align}
Here, $\delta_{i,j}^\mathrm{loop}=0$ $(1)$ when sites $i$ and $j$ belong to the same loop (different loops) in the transposition-graph. Similarly, $\delta_{h,i}^\mathrm{sublatt}$ denote whether sites $i$ and $j$ belong to the same sublattice. In fact, $T_{ii\sigma}$ is the occupation number $n_{i\sigma}$, which can be used to calculate the hole density $n_i^h$:
\begin{align}
  \label{eq:hole_density}
  n_i^h = 1 - \sum_{\sigma} n_{i\sigma} = 1 - \sum_{\sigma} \langle T_{ii\sigma}(v,v') \rangle_{v,v'} = 1 - \sum_{h\neq i} |\varphi_h(h)|^2 \left[ \delta_{h,i}^\mathrm{loop} + (1-\delta_{h,i}^\mathrm{loop}) \right] = |\varphi_h(i)|^2.
\end{align}
If we use the normalization $\sum_i |\varphi_h(i)|^2=2$ as in the main body of the paper, then the above result is $\frac{1}{2}|\varphi_h(i)|^2$. We conclude the hole density is simply $|\varphi_h(i)|^2$, which can be easily calculated without VMC.

On the other hand, for $i\neq j$, we have
\begin{align}\nonumber
  T_{ij\sigma}(v,v') &=
   \frac{1}{2^{N_{v,v'}^\mathrm{loop}-1}} \sum_{h,h'} \sum_{\sigma_h,\sigma'_{h'}} \delta_{v,\sigma_h} \delta_{v',\sigma'_{h'}} \cdot \varphi_h^\ast(h') \varphi_h(h) \Lambda^\ast(h',\sigma'_{h'}) \Lambda(h,\sigma_{h}) \cdot \eta_{\sigma_h}\eta_{\sigma'_{h'}} \\\nonumber
  &\quad\cdot
    \langle 0| c_{N\sigma'_N}\cdots c_{1\sigma'_1} c_{h'\uparrow}^\dagger (c_{i\sigma}^\dagger c_{j\sigma}) c_{h\uparrow} c_{1\sigma'_1}^\dagger\cdots c_{N\sigma'_N}^\dagger |0\rangle \\\nonumber
  &=
    \frac{1}{2^{N_{v,v'}^\mathrm{loop}-1}} \sum_{\sigma_i,\sigma'_j} \delta_{v,\sigma_i} \delta_{v',\sigma'_{j}} \cdot \varphi_h^\ast(j) \varphi_h(i) \Lambda^\ast(j,\sigma'_{j}) \Lambda(i,\sigma_{i}) \cdot \eta_{\sigma_i}\eta_{\sigma'_{j}} \\ 
  &\quad
    \cdot(-1) ( \delta_{\sigma,\uparrow} \delta_{\sigma_i(j),\uparrow} \delta_{\sigma_j(i),\uparrow} + \delta_{\sigma,\downarrow}  \delta_{\sigma_i(j),\downarrow} \delta_{\sigma_j(i),\downarrow} ) \cdot \prod_{l\neq i,j} \delta_{\sigma_i(l),\sigma_j(l)}.
\end{align}
The minus sign in the last line is the fermion sign which comes from the permutation of the $c$ and $c^\dagger$ operators. There are $2^3=8$ different cases in which the final result of $E^t_{ij\sigma}(v,v')$ takes different forms. They are classified according to: (1) the spin $\sigma$; (2) whether sites $i$ and $j$ belong to same sublattices; (3) whether sites $i$ and $j$ belong to the same loop in the transposition-graph of dimer covers $v$ and $v'$. These results are summarized in Table~\ref{tab:Et_ijs}. Note that the signs in front of the results are combinations of the fermion signs and the Marshall signs. The factor $1/2$ stems from the fact sites $i$ and $j$ belong to different loops.

\begin{table}[h]
\caption{\label{tab:Et_ijs}
Summary of $T_{ij\sigma}(v,v')$ ($i\neq j$) in $2^3=8$ different cases. The phase differences $\Delta\Lambda_{ij\sigma}^T$ are given by Eqs.~(\ref{eq:dL})-(\ref{eq:dL3}).}
\begin{center}
\begin{tabular}{|c|c|c|c|c|}
  \hline
    cases & $\sigma$ & sublattices & loops & $T_{ij\sigma}(v,v')$ \\
  \hline
  \hline
    1 & $\uparrow$ & different & same & 0 \\
  \hline
    2 & $\uparrow$ & different & different & $-\frac{1}{2} \varphi_h^\ast(j) \varphi_h(i)\, \Delta\Lambda_{ij\sigma}^{T,2}$ \\
  \hline
    3 & $\uparrow$ & same & same & $- \varphi_h^\ast(j) \varphi_h(i)\, \Delta\Lambda_{ij\sigma}^{T,3}$ \\
  \hline
    4 & $\uparrow$ & same & different & $-\frac{1}{2} \varphi_h^\ast(j) \varphi_h(i)\, \Delta\Lambda_{ij\sigma}^{T,4}$ \\
  \hline
    5 & $\downarrow$ & different & same & $\varphi_h^\ast(j) \varphi_h(i)\, \Delta\Lambda_{ij\sigma}^{T,5}$ \\
  \hline
    6 & $\downarrow$ & different & different & 0 \\
  \hline
    7 & $\downarrow$ & same & same & $- \varphi_h^\ast(j) \varphi_h(i)\, \Delta\Lambda_{ij\sigma}^{T,7}$ \\
  \hline
    8 & $\downarrow$ & same & different & 0 \\
  \hline
\end{tabular}
\end{center}
\end{table}

All the phase difference $\Delta\Lambda_{ij\sigma}^T$ in the last column of Table~\ref{tab:Et_ijs} can be divided into three parts:
\begin{align}\label{eq:dL}
  \Delta\Lambda = \Delta\Lambda_1 \cdot \Delta\Lambda_2 \cdot \Delta\Lambda_3.
\end{align}
The expressions for the above three phase difference parts are:

(i) $\Delta\Lambda_1$ comes from terms $l=i$ or $l=j$:
\begin{align}\label{eq:dL1}
  \Delta\Lambda_1 = \Lambda^\ast(j,i,\sigma) \Lambda(i,j,\sigma) = \delta_{\sigma,\uparrow} - \delta_{\sigma,\downarrow}.
\end{align}

(ii) $\Delta\Lambda_2$ comes from terms where site $l$ ($l \neq i,j$) belongs to the same loop as site $i$ or $j$ in the transposition-graph of $v$ and $v'$:
\begin{align}
  \Delta\Lambda_2 = \prod_{l\in L_i\, \mathrm{or}\, L_j,\ l\neq i,j} \Lambda^\ast(j,l,\sigma_i(l)) \Lambda(i,l,\sigma_i(l)). 
\end{align}
Note that the spin configuration on site $l$ ($l\in L_i$ or $L_j$) is totally fixed in each of the eight cases (for instance, see Fig.~\ref{fig:Ht_1loop} for case-5, and Fig.~\ref{fig:Ht_2loop} for case-2).
If we choose $\Lambda(h,l,\sigma)=e^{i\theta_{hl} \delta_{\sigma\downarrow}}$ for the ladder system, the phase difference becomes
\begin{align}
  \left. \Delta\Lambda_2 \right|_\mathrm{ladder} =
    \prod_{l\in L_i\, \mathrm{or}\, L_j,\ l\neq i,j}
    e^{i(\theta_{il}-\theta_{jl}) \delta_{\sigma_i(l),\downarrow}}.
\end{align}

(iii) $\Delta\Lambda_3$ comes from terms where site $l$ belongs to different loops as site $i$ or $j$ in the transposition-graph of $v$ and $v'$. The spin configurations on these loops have two possibilities.
\begin{align}
  \Delta\Lambda_3 = \prod_{L\neq L_i,L_j}\frac{1}{2}\sum_{\sigma_i(L)=\pm} \prod_{l\in L} \Lambda^\ast(j,l,\sigma_i(l)) \Lambda(i,l,\sigma_i(l)). 
\end{align}
Similarly, for the ladder system, this phase difference is given by
\begin{align}\label{eq:dL3}
  \left. \Delta\Lambda_3 \right|_\mathrm{ladder} = \prod_{L\neq L_i,L_j} \frac{1}{2}
  \left(\prod_{l\in L,\,l\in A}
  +\prod_{l\in L,\,l\in B}\right)
  e^{i(\theta_{il}-\theta_{jl})}.
\end{align}
The spin configuration on each loop $L$ ($L\neq L_i,L_j$) has two possibilities ($\sigma_i(L)=\pm$). $\Delta\Lambda_3$ on each loop is obtained by averaging the phase differences of these two possibilities.

The hopping energy calculation in Sec.~\ref{Appen:MC_Ht} can be viewed as special cases in Table~\ref{tab:Et_ijs}. The up-spin hopping on nearest bond corresponds to cases 1 and 2 in this table. Since case-1 has zero result, only case-2 contributes to $T_{ij\sigma}$. This is exactly Eq.~(\ref{eq:Etup}) if $-\alpha_{ij}t$ is added. Similarly, for down-spin hopping, case-5 gives the result Eq.~(\ref{eq:Etdown}), while case-6 has no contribution.

To sum up, the momentum distribution $n_\mathbf k$ is calculated from $T_{ij\sigma}$ Eq.~(\ref{eq:Tijs}). The eight cases of $T_{ij\sigma}(v,v')$ are summarized in Table~\ref{tab:Et_ijs}, with the phase differences $\Delta\Lambda_{ij\sigma}^T$ given by Eqs.~(\ref{eq:dL})-(\ref{eq:dL3}).

\subsection{Quasiparticle weight}
\label{Appen:Zk}

The quasiparticle $Z_\mathbf k$ is defined by $|\langle \mathrm{RVB} | c^\dagger_{\mathrm k\uparrow} |\Psi_G \rangle|^2$ with normalized $|\Psi_G\rangle$ and $|\mathrm{RVB}\rangle$, or equivalently
\begin{align}
  \label{eq:Zk2}\nonumber
  Z_\mathbf k &= \frac{|\langle \mathrm{RVB} | c^\dagger_{\mathrm k\uparrow} |\Psi_G \rangle|^2} {\langle\mathrm{RVB}|\mathrm{RVB}\rangle \langle\Psi_G|\Psi_G\rangle} 
  = 2 \left|\frac{1}{\sqrt{N}} \sum_i e^{-i \bold k \cdot \bold R_i} \frac{\langle \mathrm{RVB} | c^\dagger_{i\uparrow} | \Psi_G \rangle} {\langle\mathrm{RVB}|\mathrm{RVB}\rangle} \right|^2\\
  &= 2 \left| \frac{1}{\sqrt{N}} \sum_i e^{-i \bold k \cdot \bold R_i} z_i \right|^2,
\end{align}
where the normalization relation Eq.~(\ref{eq:normrel}) is used. $z_i$ is roughly the average of phase string factor $e^{-i\hat\Omega_i}$ and defined by
\begin{align}
  \label{}
  z_i &= \varphi_h(i) \frac{\langle \mathrm{RVB} | c^\dagger_{i\uparrow} e^{-i\hat\Omega_i} c_{i\uparrow}| \mathrm{RVB} \rangle }{\langle \mathrm{RVB} | \mathrm{RVB} \rangle}\\
  &=  \frac{\left(\sum_{v,v',\sigma^0}\delta_{v,\sigma_h^0} \delta_{v',\sigma_h^0}\right) w_{v'}w_v\cdot z_i(v,v')} {\left(\sum_{v,v',\sigma^0}\delta_{v,\sigma^0} \delta_{v',\sigma^0}\right) w_{v'}w_v},
\end{align}
where
\begin{align}
  \label{}
  z_i(v,v') = \varphi_h(i) \left(\frac{1}{2} \prod_{l\in L_i,\, l\neq i} \Lambda(i,l,\sigma(l))\right) \left(\prod_{L\neq L_i} \frac{1}{2} \sum_{\sigma(L)=\pm} \prod_{l\in L} \Lambda(i,l,\sigma(l))\right).
\end{align}

Similar to the VMC simulation for $|\mathrm{RVB}\rangle$, $z_i$ is obtained by averaging $z_i(v,v')$ with respect to $(v,v',\sigma^0)$ with weight $w_{v'}w_v$. $Z_\mathbf k$ is then calculated directly by using Eq.~(\ref{eq:Zk2}).

\section{Sign structure of the $\sigma$$\cdot$$t$-$J$ model}
\label{Appen:sigma}

In this appendix, we will show explicitly that the sign structure of the $\sigma$$\cdot$$t$-$J$ model, on a bipartite lattice in arbitrary dimension and hole concentration, is the Marshall sign \cite{Marshall1955}, instead of the phase string for the $t$-$J$ model. In particular, the Bloch-like wave function $|\mathrm{\bf k}_0 \rangle_{\text {BL}}$ with $\mathrm{\bf k}_0=(0,0)$ satisfies the sign structure requirement.

Let us start with a generic single-hole-doped wave function which is denoted as
\begin{align}
  \label{eq:sigma_wf}
  |\Psi\rangle = \sum_{i,\{\sigma\}} \varphi(i,\{\sigma\})\ |i,\{\sigma\}\rangle.
\end{align}
Here the basis state is defined as
\begin{align}
  \label{eq:Marshall}
  |i,\{\sigma\}\rangle \equiv c_{i\downarrow} |\{\sigma\}\rangle,
\end{align}
where the half-filled Marshall basis is given by
\begin{align}
  \label{eq:Marshall2}
  |\{\sigma\}\rangle = (-1)^{N_\downarrow^B} c_{1\sigma_1}^\dagger c_{2\sigma_2}^\dagger \cdots c_{N\sigma_N}^\dagger |0\rangle,
\end{align}
with $N_\downarrow^B$ the number of down spins belonging to sublattice $B$.

The sign structure is determined by the off-diagonal elements of the $\sigma$$\cdot$$t$-$J$ Hamiltonian. Specifically, The nonzero off-diagonal elements of the hopping terms of the $\sigma$$\cdot$$t$-$J$ model in the basis Eq.~(\ref{eq:Marshall}) are
\begin{align}
  \label{eq:tup}\nonumber
  -t\, \langle j,\{\sigma'\}| c_{i\uparrow}^\dagger c_{j\uparrow} |i,\{\sigma\}\rangle 
  &= -t\,\langle\{\sigma'\}| c_{j\downarrow}^\dagger c_{i\uparrow}^\dagger c_{j\uparrow} c_{i\downarrow} |\{\sigma\}\rangle \\
  &= t\, \langle\{\sigma'\}| S_i^+ S_j^- |\{\sigma\}\rangle \le 0,\\\nonumber
  \label{eq:tdown}
  t\,\langle j,\{\sigma'\}| c_{i\downarrow}^\dagger c_{j\downarrow} |i,\{\sigma\}\rangle
  &= t\,\langle\{\sigma'\}| c_{j\downarrow}^\dagger c_{i\downarrow}^\dagger c_{j\downarrow} c_{i\downarrow} |\{\sigma\}\rangle \\
  &= -t\, \langle\{\sigma'\}| n_i n_j |\{\sigma\}\rangle \le 0.
\end{align}
On the other hand, the nonzero off-diagonal elements of the superexchange terms are
\begin{align}
  \label{eq:J}
  \frac{J}{2}\,\langle h,\{\sigma'\}| S_i^+ S_j^- |h,\{\sigma\}\rangle \le 0.
\end{align}
The nonnegativity of both Eq.~(\ref{eq:tup}) and Eq.~(\ref{eq:J}) is owing to the Marshall sign $(-1)^{N_\downarrow^B}$ in the basis Eqs.~(\ref{eq:Marshall}) and (\ref{eq:Marshall2}). We conclude the off-diagonal elements of the $\sigma$$\cdot$$t$-$J$ Hamiltonian are all non-positive in the basis Eq.~(\ref{eq:Marshall}).

As a result, according to the Perron-Frobenius theorem, the ground state of the $\sigma$$\cdot$$t$-$J$ model has the form of Eq.~(\ref{eq:sigma_wf}) with $\varphi(i,\{\sigma\}) \ge 0$. That means, the sign structure of the $\sigma$$\cdot$$t$-$J$ model is exactly the Marshall sign \cite{Marshall1955} in the basis Eq.~(\ref{eq:Marshall2}), the same as the Heisenberg spin model. In particular, if we ignore the spin polaron effect, the Bloch-like wave function $|\mathrm{\bf k}_0 \rangle_{\text {BL}}$ at $\mathrm{\bf k}_0=(0,0)$ should well describe the $\sigma$$\cdot$$t$-$J$ model, for it satisfies the sign structure of this model. Indeed, Fig.~\ref{fig:stJ} illustrates the overall agreement of the Bloch state $|\mathrm{\bf k}_0 \rangle_{\text {BL}}$ at $\mathrm{\bf k}_0=(0,0)$ with the DMRG result.

\end{document}